\documentclass[fleqn,usenatbib,twocolumn]{mnras}
\usepackage[T1]{fontenc}

\DeclareRobustCommand{\VAN}[3]{#2}
\let\VANthebibliography\thebibliography
\def\thebibliography{\DeclareRobustCommand{\VAN}[3]{##3}\VANthebibliography}

\usepackage{array}
\usepackage{graphicx}  % Including figure files
\usepackage{makecell}
\usepackage{adjustbox}

\usepackage{amsmath}   % Advanced maths commands
\usepackage[left]{lineno}
\usepackage{epstopdf}
\IfFileExists{ulem.sty}{\usepackage{ulem}}{}
\usepackage{amssymb}
\usepackage{times}
\IfFileExists{pstricks.sty}{\usepackage[usenames,dvipsnames]{pstricks}}{}
\usepackage{psfrag}
\usepackage{lipsum}
\usepackage{natbib}
\usepackage{rotating}
\usepackage{hyperref}
\ifdefined\pdfminorversion\pdfminorversion=7\fi
\makeatletter
\def\caption{\H@refstepcounter\@captype\@dblarg{\@caption\@captype}}
\makeatother
\usepackage{booktabs}
\IfFileExists{newtxtext.sty}{\usepackage{newtxtext,newtxmath}}{\usepackage{mathptmx}}
\usepackage{orcidlink}

\def\apjl{ApJ}%% Astrophysical Journal, Letters
\def\apjs{ApJS}%% Astrophysical Journal, Supplement
\def\ao{Appl.~Opt.}%% Applied Optics
\def\aap{A\&A}%% Astronomy and Astrophysics
\newcommand{\be}{\begin{equation}}
\newcommand{\ee}{\end{equation}}
\newcommand{\bary}{\begin{eqnarray}}
\newcommand{\eary}{\end{eqnarray}}

\title[GRB 260924A: Extreme Energetics and Broadband Emission]{
GRB 260924A: Extreme Energetics, Bright GeV Emission, and Exceptional X-ray and Optical Luminosity}

\author[Fraija, N.]{
Nissim~Fraija\,\orcidlink{0000-0002-0173-6453},$^{1}$\thanks{E-mail: nifraija@astro.unam.mx} Peter~Veres\,\orcidlink{0000-0002-2149-9846},$^{2}$
Rosa~L.~Becerra\,\orcidlink{0000-0002-0216-3415},$^{1}$ Antonio~Galván-Gámez\,\orcidlink{0000-0001-5193-3693},$^{1}$ Thomas Freeman\,\orcidlink{0009-0005-4754-8836},$^{2}$
\newauthor
Simone~Dichiara\,\orcidlink{0000-0001-6849-1270},$^{3}$ and
Maria~G.~Dainotti\,\orcidlink{0000-0003-4442-8546}$^{4,5}$
\\
$^{1}$Instituto de Astronom\'ia, Universidad Nacional Aut\'{o}noma de M\'{e}xico, A. P. 70-264, Cd. Universitaria, Ciudad de M\'{e}xico, M\'{e}xico\\
$^{2}$Center for Space Plasma and Aeronomic Research (CSPAR), University of Alabama in Huntsville, Huntsville, AL 35899, USA\\
$^{3}$Department of Astronomy and Astrophysics, The Pennsylvania State University, 525 Davey Lab, University Park, PA 16802, USA\\
$^{4}$Division of Science, National Astronomical Observatory of Japan, 2-21-1 Osawa, Mitaka, Tokyo 181-8588, Japan\\
$^{5}$The Graduate University for Advanced Studies (SOKENDAI), 2-21-1 Osawa, Mitaka, Tokyo 181-8588, Japan\\
}
\date{Accepted XXX. Received YYY; in original form ZZZ}

\pubyear{\the\year{}}

\begin{document}
\label{firstpage}
\pagerange{\pageref{firstpage}-\pageref{lastpage}}
\maketitle
%\linenumbers

%BAT emission over 250--320~s further demonstrates that the intervening interval is not completely quiescent.

%LAT detects emission from approximately 70 to 1100~s, including the highest-energy photon associated with the burst, at 43.31~GeV and 82~s, and a 14.24~GeV photon at 304~s. Its light curve reveals an early excess above the subsequent decay.

\begin{abstract}
We present a multiwavelength study of GRB~260924A
at $z=3.51\pm0.10$ using \textit{Fermi}, \textit{Swift},
and ground-based optical and near-infrared observations.
Its isotropic-equivalent gamma-ray energy,
$E_{\gamma,\mathrm{iso}}\simeq9.79\times10^{54}$~erg,
places it among the most energetic GRBs observed.
The prompt emission comprised two bright episodes
separated by weaker activity.
The X-ray afterglow ranked among the most luminous
in our comparison sample at selected post-flare epochs,
with a luminosity comparable to that of GRB~221009A
at a rest-frame time of 3.5~ks.
At a common redshift of $z=1$, the optical afterglow
outshone GRB~221009A by approximately 2~mag at 0.3~d
and 1~mag near 1~d.
LAT emission extended to approximately 1100~s
and showed an early excess above the subsequent decay.
The highest-energy photon, 43.3~GeV, arrived at 82~s
during the first episode; a 14.2~GeV photon arrived
at 304~s between episodes.
We describe the afterglow light curves through
forward-shock synchrotron and synchrotron self-Compton
(SSC) emission, with reverse-shock SSC accounting
for the early LAT excess.
We connect the early and late afterglow phases
by modelling the X-ray rebrightening at several
kiloseconds through energy injection from trailing
ejecta and evolving shock microphysics.
The late afterglow suggests a stratified medium,
modelled as $n(R)\propto R^{-1.2}$.
Within this interpretation, the magnetic contrast
$R_B\equiv(\varepsilon_{B,r}/\varepsilon_{B,f})^{1/2}
\simeq2.6$ suggests magnetized ejecta,
while the highest-energy LAT photons exceed
the predicted synchrotron limit, supporting
an SSC contribution.
\end{abstract}


%%%%%%%%%%%%%%%%%%%%%%%%%%%%%%%%%%%%%%%%%%%%%%%%%%

%%%%%%%%%%%%%%%%% BODY OF PAPER %%%%%%%%%%%%%%%%%%

\section{Introduction}
\label{sec:introduction}

Gamma-ray bursts (GRBs) are high-energy transients
whose prompt emission exhibits diverse temporal
structures, with rapid variability and durations
ranging from milliseconds to minutes
\citep{vonKienlin2020}. Their emission can continue well beyond the brightest gamma-ray pulses, motivating observations across multiple energy bands and timescales \citep{Kumar2015}.

Relativistic outflows provide a framework for
interpreting GRB emission. The prompt emission
may arise from internal dissipation within the
outflow \citep{1994ApJ...430L..93R} and radiation
released near its photosphere
\citep{2000ApJ...530..292M}.
As the ejecta interact with the surrounding medium,
a forward shock propagates into it and a reverse
shock can cross the ejecta
\citep{1997ApJ...476..232M,2000ApJ...545..807K}.
Electrons accelerated at these shocks may produce
synchrotron emission \citep{1998ApJ...497L..17S},
while inverse-Compton scattering can contribute
at higher photon energies
\citep{2001ApJ...548..787S}.
The relative contributions of these processes
depend on the outflow properties and environment
and can be constrained by the observed temporal
and spectral evolution.

The combination of \textit{Fermi} \citep{Meegan2009,Atwood2009} and \textit{Swift} \citep{Gehrels2004} provides a broad view of GRB emission from the prompt phase to the developing afterglow. The \textit{Fermi} Gamma-ray Burst Monitor (GBM) and the \textit{Swift} Burst Alert Telescope (BAT) detect and characterize the temporal and spectral structure of the prompt gamma-ray emission, while the \textit{Fermi} Large Area Telescope (LAT) extends this view to GeV energies. LAT observations have revealed that this high-energy emission can show a delayed onset and persist well beyond the brightest lower-energy gamma-ray pulses \citep{Ajello2019LATcatalog}. Following the trigger, the rapid response of \textit{Swift} provides a complementary view at
lower energies, with the X-Ray Telescope
(XRT; \citealt{2005SSRv..120..165B}) tracing the
early X-ray evolution and the Ultraviolet/Optical
Telescope (UVOT; \citealt{2005SSv..120...95R})
identifying and monitoring the ultraviolet (UV)
and optical counterpart.  Ground-based optical and near-infrared (NIR) observations extend this multiwavelength coverage.

Multiwavelength observations provide complementary
constraints on the radiation mechanisms and emitting
regions responsible for the temporal and spectral
structure of GRBs \citep{2015ApJ...804..105F, 2016ApJ...818..190F, Ajello2019LATcatalog, 2019ApJ...883..162F, Becerra2023, Dainotti2024}. The combined afterglow evolution across different
bands probes the properties of the relativistic
ejecta and the density structure of the circumburst
medium \citep{2017ApJ...848...15F}.
Departures from a simple decay can also reveal
the need to consider additional emission components,
continued energy supply \citep[e.g., see][]{2015ApJ...814....1L, 2021ApJ...906...60Z, 2019ApJ...872..118B, 2022MNRAS.511.6205P, 2021ApJ...918...12F}, or evolving shock
microphysics \citep[e.g., see][]{2003MNRAS.339..881R, 2013MNRAS.428..845L, 2024MNRAS.527.1884F}.  Broad spectral and temporal coverage therefore
allows afterglow models to be tested against
several observational constraints within a common
physical framework.

GRB 260924A was detected by both \textit{Fermi} \citep{GCN45712} and \textit{Swift} \citep{GCN45713}. Its GBM and BAT light curves show a bright, structured initial episode followed by a second
prominent gamma-ray pulse beginning near 334~s after the GBM trigger \citep{GCN45733}. Between these episodes, BAT detects weaker emission, while \textit{Fermi}/LAT reveals high-energy activity extending to tens of GeV \citep{GCN45721}. XRT observed the decline following the initial activity and the rise associated with the second pulse, although its early coverage ended before that episode was complete.   \textit{Swift}/UVOT and ground-based observations provided
optical and NIR follow-up, complementing
the XRT coverage and enabling multiwavelength
monitoring over the following days. Optical spectroscopy revealed both the Ly$\alpha$ break and the Lyman limit, yielding a redshift of $z=3.51\pm0.10$ \citep{GCN45740}.   Using $z=3.51$ and the time-integrated GBM spectrum, \citet{GCN45750} estimated an isotropic-equivalent gamma-ray energy of $E_{\gamma,\mathrm{iso}}\simeq9.74\times10^{54}\,\mathrm{erg}$ in the 1--10,000~keV rest-frame band, placing GRB 260924A among the most energetic GRBs detected and at an energy scale comparable to that of GRB 221009A \citep[see e.g.][]{Burns2023,OConnor2023,Lesage2023}.  The combination of a large prompt energy release, extended GeV activity, and extensive afterglow coverage makes GRB~260924A a useful case for examining the relationship between the prompt
emission and the subsequent broadband evolution.

%Comparison with the optical afterglow population \citep{Oates2009,Kann2011,Dainotti2024,Becerra2023} places GRB 260924A among the most luminous optical afterglows observed.

In this paper, we present a multiwavelength analysis
of GRB~260924A, combining \textit{Fermi}/GBM and LAT
observations with \textit{Swift}/BAT, XRT, and UVOT
data and ground-based optical and NIR
follow-up. We characterize the temporal and spectral evolution
of the two bright gamma-ray episodes, the intervening
activity, and the subsequent afterglow.  We use the light curves, spectral measurements,
and LAT photon energies to assess possible radiation
mechanisms and emitting regions. We model the afterglow with forward-shock synchrotron and synchrotron self-Compton (SSC)  emission, an early reverse-shock SSC contribution, and energy injection with evolving shock microphysics. We also place the event in the context of GRB
comparison samples using rest-frame X-ray
luminosities, optical light curves transformed
to a common redshift, and observed LAT photon fluxes.

The paper is organized as follows. Section~\ref{sec:observations} describes the observations, data analysis, and common temporal
reference. Section~\ref{sec:temporal} presents the temporal evolution across the observed energy bands. Section~\ref{sec:spectra} presents
the BAT, GBM, and LAT spectral analyses and selected
LAT high-energy events.  Section~\ref{sec:description} presents our external-shock model of the multiwavelength afterglow and the inferred physical parameters. Section~\ref{sec:discussion} discusses the physical implications of these results, and Section~\ref{sec:conclusions} summarizes our conclusions.

\section{Observations and Data Analysis}
\label{sec:observations}

This section describes the observations and analysis
procedures used to construct the multiwavelength
data set of GRB~260924A, including the processing
of the \textit{Fermi} and \textit{Swift} data and
the compilation of ground-based optical and NIR
photometry. Unless otherwise stated, times are
given in the observer frame relative to the
\textit{Fermi}/GBM trigger, adopted as $T_0$.

\subsection{\textit{Fermi}/GBM}
\label{sec:gbm_analysis}

\subsubsection{Light curves}
\begin{figure*}
\centering
\includegraphics[width=0.5\linewidth]{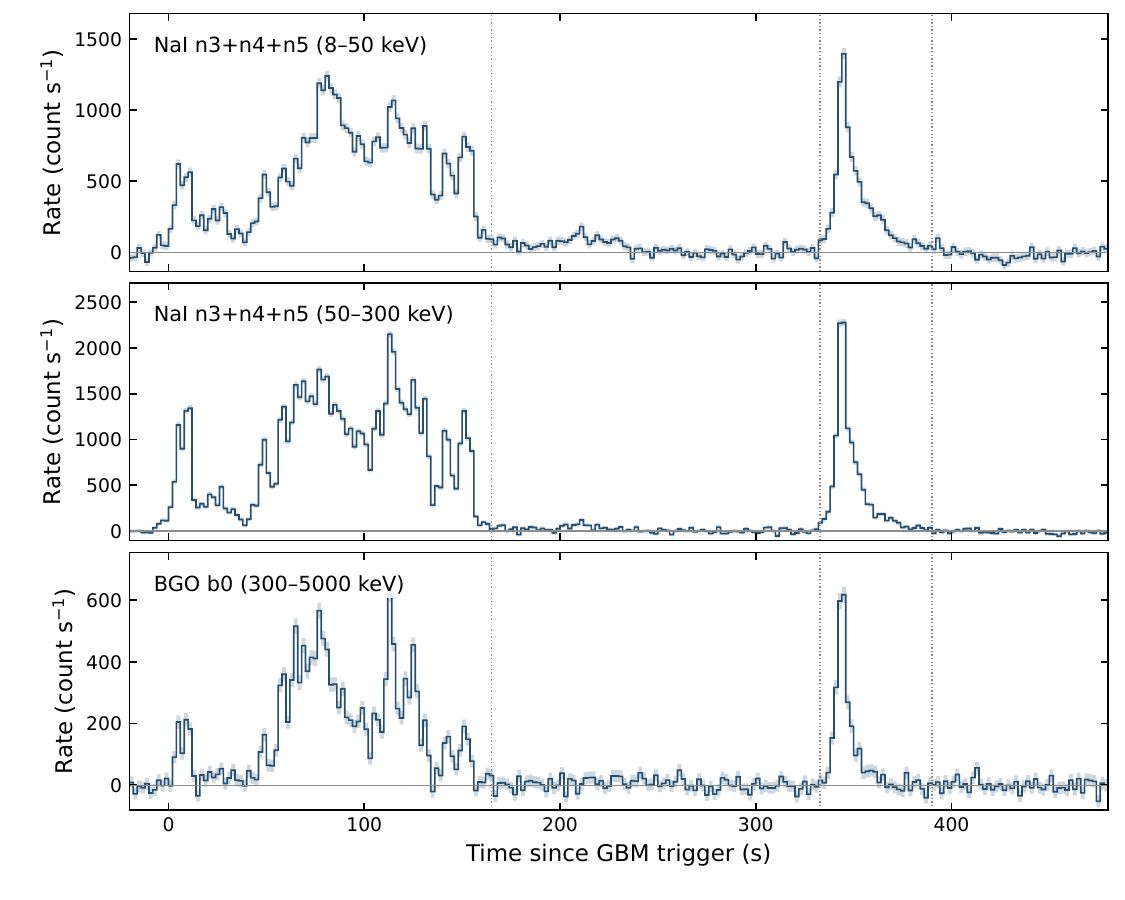}
\caption{Background-subtracted \textit{Fermi}/GBM light curves
of GRB~260924A in 2-s bins. The upper panels show
the summed NaI n3, n4, and n5 count rates in
8--50 and 50--300~keV; the lower panel shows
BGO b0 in 300--5000~keV.
Shaded envelopes represent nominal $1\sigma$
uncertainties from counting statistics and the
fitted linear background.
Vertical dotted lines mark the adopted temporal
boundaries at 164.9, 332.8, and 390~s.
All times are relative to the GBM trigger.}
\label{fig:gbm_lightcurve}
\end{figure*}

GRB~260924A was detected by the {\itshape Fermi}/GBM instrument at  2026 September 24 01:13:23 UTC ($T_0$) \citep{GCN45712}. We constructed the GBM count-rate light curves directly from time-tagged event (TTE) data in
2-s bins relative to $T_0$ (Figure~\ref{fig:gbm_lightcurve}). We selected PHA channels whose energy boundaries lay entirely within the nominal bands of
8--50 and 50--300~keV for NaI and 300--5000~keV for BGO. Count rates were corrected for detector deadtime. For each detector and band, a linear background
was fitted over $[-125,-35]$ and $[250,320]$~s relative to $T_0$ and subtracted.
The resulting rates from NaI detectors n3, n4, and n5 were summed within each band, while BGO b0 was displayed separately. Uncertainties include counting statistics and the covariance of the fitted background parameters.

Although automated methods such as Bayesian blocks
can be used to define time intervals, we selected
the boundaries by visual inspection of the GBM
light curves to delimit the main emission episodes
and facilitate comparison with BAT and LAT.
The adopted boundaries are marked in
Figure~\ref{fig:gbm_lightcurve}.

\subsubsection{Spectral analysis}

For the first bright episode, we analysed the spectrum integrated over $-6.1$--$164.9$~s and the three subintervals $-6.1$--$50$, $50$--$115$, and $115$--$164.9$~s relative to $T_0$. We used \ion{Na}{I} detectors n3, n4, and n5 and \ion{BGO}{}
detector b0, with nominal energy ranges of
8--900~keV and 0.3--40~MeV, respectively.  

For each detector, we constructed an effective response by averaging the time-dependent RSP2 response over the corresponding interval, weighting by the total observed counts.

The background uncertainties, including the covariance of the fitted coefficients, were propagated to the integrated background counts. Further details of the response construction, background estimation, and exposure calculation are given in Appendix~\ref{app:gbm_systematics}.

Spectral fitting used a Poisson--Gaussian profile likelihood, combining Poisson-distributed observed counts with Gaussian constraints on the estimated background counts. We report the fit statistic as ${\rm PG}=2S$, where $S$ is the negative log-likelihood up to data-dependent constants. We modelled the continuum using either the Band function \citep{Band1993}, which smoothly joins a low-energy power law with an exponential cutoff to a high-energy power law, or the Comptonized model, defined here as a power law with an exponential cutoff.  We also tested whether adding a blackbody (BB) improved the Band or Comptonized fits
\citep{Ryde2004,RydePeer2009}. The BB prescription and parameter bounds are given in Appendix~\ref{app:gbm_blackbody}.

To examine the earliest spectral evolution,
we divided the $-6.1$ to $50$~s interval at
$12.29$~s, selecting this boundary from the
0.512-s GBM light curve before fitting the spectra.
The second bright episode was analysed over
$332.8$--$390$~s, with $334$--$390$~s used
as an alternative extraction interval.

Spectral fits were performed in detector count space;
unfolded $\nu F_\nu$ spectra are shown for visualization.
Tests of sensitivity to the response treatment,
background model, and channel selection are
described in Appendix~\ref{app:gbm_systematics}.

The spectra displayed in Figures~\ref{fig:gbm_spectra} and~\ref{fig:gbm_spectra_eps1} and the parameters listed in Tables~\ref{tab:gbm_spectral_parameters} and~\ref{tab:gbm_spectral_normalizations} use linear
temporal backgrounds; quadratic backgrounds were
examined as a sensitivity test.

\subsection{\textit{Fermi}/LAT}
\label{sec:lat_analysis}

We analysed \textit{Fermi}/LAT events within a $15^{\circ}$ region of interest (ROI) centred on the \textit{Swift}/XRT position, $(\mathrm{RA},\mathrm{Dec})=(133.76^{\circ},12.04^{\circ})$, over the 0.1--1000~GeV energy range. We selected event class~8 and event type~3, applied a maximum zenith angle of $100^{\circ}$, and used the \texttt{P8R3\_TRANSIENT020E\_V3} instrument response functions. 
We used an unbinned maximum-likelihood analysis with
{\sc Fermitools} version 2.5.3 and \texttt{fermitools-data} version 0.18.

We analysed five non-overlapping intervals spanning
$T_0+70.04$ to $T_0+1100$~s (Table~\ref{tab:lat_integrated_spectra}), covering
the initial episode, inter-episode activity, the
second pulse, and the subsequent high-energy tail.
Alternative overlapping intervals were used to
test sensitivity to the temporal selection and
were not included as independent light-curve
measurements (Section~\ref{sec:lat_lightcurve}).

The background model included the Galactic diffuse template \texttt{gll\_iem\_v07.fits}, the isotropic component \texttt{iso\_P8R3\_TRANSIENT020E\_V3\_v1.txt}, and 243 sources from the 4FGL-DR4 catalogue within $25^{\circ}$ of the \textit{Swift}/XRT position. The same catalogue sources were included in all five principal intervals and were modelled as point sources with their positions and spectral parameters fixed.   The GRB spectrum was described by a power law, $dN/dE\propto E^{-\Gamma_{\rm LAT}}$, where $\Gamma_{\rm LAT}$ is
the photon index.  Under the source hypothesis, the free parameters were the GRB normalization and photon index and the normalizations of the Galactic and isotropic diffuse components. The GRB position and diffuse spectral shapes were held fixed. All events satisfying the selection criteria were included in the likelihood analysis without an additional cut on their fitted source-association probabilities.

For each interval, we evaluated the likelihood-ratio
test statistic (TS) by comparing models with and
without the GRB, as described in Appendix~\ref{app:lat_systematics}.
Both diffuse normalizations were reoptimized under each hypothesis within the bounds $[0,30]$.  The quoted photon-flux and photon-index errors are nominal $1\sigma$ statistical uncertainties derived from the local covariance matrix.

Energy fluxes were obtained by integrating the
energy-weighted fitted spectra over 0.1--1000~GeV.
Their uncertainties were propagated from the
fitted covariance matrix, including the covariance
between normalization and photon index. For the light-curve display in Figure~\ref{fig:LAT_lightcurve}, we
integrated the same fitted models over
0.1--300~GeV; the likelihood fits, Table~\ref{tab:lat_integrated_spectra}, and temporal analysis use 0.1--1000~GeV.

We also examined the energies and arrival times of individual LAT events. Their model-dependent source-association probabilities,
$P_{\rm GRB}$, were obtained from a fit over 0--1100~s after the GBM trigger (hereafter the full-window fit). The bottom-left panel of Figure~\ref{fig:LAT_lightcurve}  displays all 404 selected events, with greyscale indicating these probabilities. Table~\ref{tab:lat_high_energy_events} lists the ten highest-energy events with $P_{\rm GRB}\geq0.9$. Their properties are discussed in Section~\ref{sec:lat_photons}.

Further implementation details and sensitivity
tests are provided in Appendix~\ref{app:lat_systematics}.

\subsection{\textit{Swift}/BAT}
\label{sec:swift}

\textit{Swift}/BAT triggered on and located GRB~260924A \citep{GCN45713}. Throughout this work, \textit{Swift} times were converted to the \textit{Fermi}/GBM reference time using
$t_{\rm GBM}=t_{\rm BAT}+4.71$~s. 

The BAT count-rate light curve was constructed
from mask-weighted rates in the 15--350~keV band
using 1-s bins and is shown in the upper panel
of Figure~\ref{fig:BAT_photon}.   It was used to compare the temporal structure with that observed by GBM and XRT (Section~\ref{sec:second_pulse}). Spectral fits
were performed over 15--150~keV.

We extracted BAT spectra over $T_0+70$ to
$T_0+165$~s, $T_0+250$ to $T_0+320$~s,
and $T_0+334$ to $T_0+390$~s. These intervals
sample the initial bright episode, the weaker
inter-episode activity, and the second bright
pulse, respectively. We fitted each spectrum
with a power-law model and examined whether
adding an exponential cutoff improved the fit.
The spectral responses accounted for changes
in spacecraft pointing within each interval.
Table~\ref{tab:bat_pl} summarizes the power-law fits.

To examine spectral evolution within the two
bright episodes, we also extracted spectra in
shorter subintervals and fitted them with
power-law models. The resulting photon indices,
$\Gamma_{\rm BAT}$, are shown in the lower panels
of Figure~\ref{fig:BAT_photon} and discussed in Section~\ref{sec:bat_spectra}.

The spectra were fitted using the $\chi^2$
statistic, including channel-dependent systematic
uncertainties. We report 90 per cent confidence
intervals for one parameter of interest,
corresponding to $\Delta\chi^2=2.706$.  Energy-flux confidence intervals were obtained by profiling over the photon index, accounting for the normalization--index covariance. Details of the treatment and verification of the BAT systematic uncertainties
are provided in Appendix~\ref{app:bat_systematics}.

\subsection{\textit{Swift}/XRT}
\label{sec:xrt}

\textit{Swift}/XRT began its pointed observations
of GRB~260924A approximately 110.6~s after the
BAT trigger, detecting a bright, previously
uncatalogued X-ray source within the BAT localization \citep{GCN45713}. Subsequent astrometric refinement yielded a position of $\mathrm{RA}=133.76^{\circ}$ and $\mathrm{Dec}=+12.04^{\circ}$ (J2000), with a 90 per cent confidence radius of $2.1''$ \citep{GCN45715}. The refined XRT analysis included 6.4~ks of observations spanning approximately 99~s to 43.5~ks after the BAT trigger, comprising 365~s in Windowed Timing (WT) mode, including the initial slew data, followed by observations in Photon Counting (PC) mode \citep{GCN45730}.

We obtained the \textit{Swift}/XRT 0.3--10~keV count-rate
and flux light curves, together with the
hardness-ratio products, from the UK Swift
Science Data Centre (UKSSDC) GRB repository \citep{Evans2009,Evans2010}\footnote{\url{https://www.swift.ac.uk/}}.  Figure~\ref{fig:XRT_lightcurve} shows the
count-rate evolution and the photon index
$\Gamma_{\rm XRT}$ inferred from the hardness
ratio. Table~\ref{tab:xrt_catalogue_temporal} reproduces the automated temporal
fit reported by the catalogue, retaining its
original BAT time reference. Times plotted in
Figure~\ref{fig:XRT_lightcurve} were converted to the GBM reference
using the offset specified in Section~\ref{sec:swift}.

The X-ray temporal analysis and comparison with
the optical/NIR afterglow are presented in
Section~\ref{sec:lc_sed}, with the late-time break test
described in Appendix~\ref{app:xrt_late_break_test}.

To characterize the early X-ray spectral evolution
outside the flare interval, we analysed three
spectra accumulated over 263.5--314.0~s (WT),
3652.5--3757.0~s (WT), and 3763.4--4656.0~s (PC)
after the GBM trigger. We fitted these spectra
jointly over 0.3--10~keV with an absorbed power-law
model. The Galactic absorbing column was fixed,
and the intrinsic column at $z=3.51$ was tied
across the three spectra, while their photon
indices and normalizations varied independently.

The fits used the W statistic, with spectra grouped to at least one source count per bin. Parameter uncertainties correspond to 90 per cent
confidence for one parameter of interest ($\Delta W=2.71$), with the remaining parameters reoptimized. Table~\ref{tab:xrt_common_absorption} summarizes the fitted parameters and absorption-corrected flux densities. Figure~\ref{fig:XRT_spectra} shows the spectra, the models folded through their instrumental responses, and the corresponding residuals. Additional regrouping was applied solely for
visualization and did not affect the spectral fits.

\subsection{UV, optical, and near-infrared observations}
\label{sec:optical_observations}

We compiled the available UV, optical, and NIR photometry of GRB~260924A reported by multiple follow-up facilities. The data set includes observations obtained with the Ultraviolet/Optical Telescope (UVOT) on board \textit{Swift} \citep{GCN45735}; the 1.5-m telescope at Sierra Nevada Observatory (OSN) \citep{GCN45717}; the Nordic Optical Telescope (NOT) with ALFOSC \citep{GCN45718}; the Lulin One-meter Telescope (LOT) and Super Light Telescope (SLT) \citep{GCN45748}; the GROWTH-India Telescope \citep{GCN45734}; the SYSU 80-cm telescope \citep{GCN45736,GCN45758}; the Ond\v{r}ejov D50 telescope \citep{GCN45743}; the Calar Alto 2.2-m telescope with CAFOS \citep{GCN45740}; the 1.3-m COLIBRÍ telescope with the two-channel DDRAGO imager \citep{GCN45759}; the MDM 1.3-m telescope with MIRAGE \citep{GCN45762}; and AbAO \citep{GCN45774}.

We placed the photometry on the AB magnitude system and corrected it for Galactic foreground extinction adopting $E(B-V)=0.041$~mag \citep{Schlafly2011} and $R_V=3.1$ \citep{Fitzpatrick1999}. For visualization, measurements in related passbands were displayed together as $g/g'$ and $r/r'/R$. The temporal fit reported in
Section~\ref{sec:lc_sed} used only the six $r/r'$ detections,
while the original filter information was retained for the spectral energy distribution analysis. Upper limits were retained at the confidence level reported by the corresponding observations.

The upper panel of Figure~\ref{fig:lc_sed} shows the compiled
UV, optical, and NIR light curves together with the XRT flux-density light curve at 1~keV. The temporal evolution and optical-to-X-ray spectral energy distributions are discussed
in Section~\ref{sec:lc_sed}.

\section{Multiwavelength temporal evolution}
\label{sec:temporal}

We characterize the multiwavelength evolution of
GRB~260924A across the two bright gamma-ray episodes,
the intervening activity, and the subsequent afterglow.
The temporal comparisons and optical-to-X-ray spectral
energy distributions (SEDs) provide the observational basis
for the modelling presented in Section~\ref{sec:description}.

\subsection{\textit{Fermi}/GBM and \textit{Swift}/BAT light curves}
\label{sec:gbm_bat_lightcurves}

Figures~\ref{fig:gbm_lightcurve} and~\ref{fig:BAT_photon}
show the GBM and BAT count-rate light curves.
The initial bright, structured GBM episode extends
to approximately $T_0+160$--$165$~s, followed by
a distinct second pulse beginning near $T_0+334$~s
that is also evident in BAT.
Between these episodes, a candidate GBM excess
at approximately $T_0+195$--$235$~s is most apparent
in the NaI detectors, while BAT also traces weaker
emission.
Count-rate amplitudes are not directly comparable
between instruments because their energy ranges
and responses differ.

Because $T_0+250$--$320$~s was used to fit the GBM
background, the background-subtracted light curve
does not provide an independent test for source
emission during that interval.

\subsection{\textit{Fermi}/LAT light curve and inter-episode emission}
\label{sec:lat_lightcurve}

Figure~\ref{fig:LAT_lightcurve} shows the LAT light
curve for the five intervals analysed in
Section~\ref{sec:lat_analysis}.
The corresponding photon and energy fluxes are
listed in Table~\ref{tab:lat_integrated_spectra}.

LAT emission was detected between the two bright
gamma-ray episodes. During $T_0+250$ to
$T_0+320$~s, the fitted source at the XRT position
has $\mathrm{TS}=87.16$, a 0.1--1000~GeV photon
flux of $(5.78\pm1.41)\times10^{-5}$
ph~cm$^{-2}$~s$^{-1}$, and
$\Gamma_{\rm LAT}=2.06\pm0.24$.

Subdividing the first principal interval at
$T_0+104$~s reveals a decrease in the average
0.1--1000~GeV photon flux from
$(11.45\pm1.01)\times10^{-4}$ over
70.04--104~s to $(3.25\pm0.39)\times10^{-4}$
ph~cm$^{-2}$~s$^{-1}$ over 104--164.9~s,
relative to $T_0$. The lower average flux in the later subinterval
indicates an early decline, although these two
measurements do not resolve the peak time.
Neither these subdivisions nor the overlapping
250--320~s interval were included as additional
independent points in the five-bin temporal fit.

The four principal intervals after
$T_0+164.9$~s are described by
$F_{\rm ph}(t)=A(t/164.9~{\rm s})^{-\alpha_{\rm LAT}}$,
with $\alpha_{\rm LAT}=1.30^{+0.20}_{-0.19}$,
$A=(1.60^{+0.29}_{-0.27})\times10^{-4}$
ph~cm$^{-2}$~s$^{-1}$, and
$\chi^2/{\rm dof}=0.31/2$.
The model was averaged over each interval using
normalized exposure weights.
Parameter uncertainties correspond to
$\Delta\chi^2=1$, with the other parameter
reoptimized.

The first interval, excluded from this fit,
has a measured photon flux of
$(6.03\pm0.44)\times10^{-4}$
ph~cm$^{-2}$~s$^{-1}$, compared with an
extrapolated interval-averaged prediction of
$(2.68\pm0.67)\times10^{-4}$
ph~cm$^{-2}$~s$^{-1}$.
Their ratio, $2.25\pm0.59$, quantifies the
early excess relative to the subsequent decay.
The fit includes the interval containing the
second bright episode, and its uncertainties
are conditional on the adopted spectral models
and exposure weights.
Further details and sensitivity tests are given
in Appendix~\ref{app:lat_temporal_weights}.

% The initial analysis reported emission above 100~MeV during the first 1.1~ks, with a photon flux of $(8.3\pm0.5)\times10^{-5}$~ph~cm$^{-2}$~s$^{-1}$ and a photon index $\Gamma_{\mathrm{LAT}}=2.2\pm0.1$  \citep{GCN45721}. 

\subsection{Temporal structure and hardness evolution of the second bright episode}
\label{sec:second_pulse}

A second prominent gamma-ray pulse begins near
$T_0+334$~s in both the GBM and BAT light curves. XRT records a contemporaneous flare, with an observed
maximum near $T_0+344.5$~s followed by a partial decline.
Its early coverage ends near $T_0+359.4$~s, preventing
us from following the complete X-ray decay.

Figure~\ref{fig:second_pulse} compares the pulse
profiles and hardness evolution across GBM, BAT,
and XRT. The GBM pulse is broader in 8--50~keV
than in 50--300~keV. The hardness measurements
are listed in Table~\ref{tab:second_pulse_hardness}.

GBM and BAT show decreasing hardness during
the well-sampled early decay following the pulse
maximum. XRT shows hardening during the rise,
but its limited coverage prevents characterization
of the spectral evolution throughout the full decay.

We fitted the 19 original WT bins covering
344.77--359.37~s, excluding the peak bin,
with a power-law flare component proportional
to $(t-332.8~{\rm s})^{-\alpha_{\rm XRT}}$.
The underlying emission was represented by
$B_{300}(t/300~{\rm s})^{-\beta}$, with its
normalization and slope jointly constrained
by 30 pre-flare bins spanning 263.50--314.04~s.
The fit yields $\alpha_{\rm XRT}=1.11\pm0.17$,
with uncertainties quoted at the $1\sigma$ level
($\Delta\chi^2=1$ for one parameter of interest).
The power-law model gives $\chi^2/{\rm dof}=28.73/45$,
whereas a constant excess above the underlying
emission gives $\chi^2/{\rm dof}=70.57/46$. A reconstruction using the time-dependent spectral conversion gives a compatible flux decay.
The temporal index depends strongly on the
assumed time origin, and the XRT hardness
is consistent with remaining constant during
the observed decline.

\subsection{UV, optical, NIR, and X-ray evolution}
\label{sec:lc_sed}
%Hereafter, we use \textit{rebrightening} for this later feature and \textit{flare} for the X-ray pulse accompanying the second bright gamma-ray episode.
%The flare associated with the second bright gamma-ray episode is distinguished from this later rebrightening.

The XRT light curve shows an initial decline interrupted
by the flare discussed in Section~\ref{sec:second_pulse}. After the flare, the observations resume with a declining light curve,
followed by a rebrightening at several kiloseconds
(Figure~\ref{fig:XRT_lightcurve}).  The automated catalogue fit describes this feature with a rising fourth segment, $\alpha_4=-1.50$, between breaks
near 4.7 and 6.3~ks (Table~\ref{tab:xrt_catalogue_temporal}). 
Hereafter, we refer to the afterglow phases before and after this rebrightening as the
\textit{early} and \textit{late afterglow}, respectively.

The upper panel of Figure~\ref{fig:lc_sed} compares
the UV, optical, and NIR light curves compiled
in Section~\ref{sec:optical_observations} with the XRT evolution at 1~keV. The observations span from a few minutes to
several days after the burst, allowing us to
characterize the late-time decay and reconstruct
the optical-to-X-ray spectral energy distribution
at selected epochs.

A single-power-law fit to six $r/r'$ detections spanning
0.19--2.44~d gives $\alpha_{\rm opt}=1.46\pm0.01$
for $F_\nu\propto t^{-\alpha_{\rm opt}}$, with
$\chi^2/{\rm dof}=25.18/4$. The excess scatter indicates
that this slope characterizes the overall decline
but does not fully describe the optical evolution.

For comparison, a power-law fit to 69 late-time XRT PC-mode measurements with bin centres spanning 0.17--3.40~d yields $\alpha_X=1.57\pm0.04$ for $F_\nu\propto t^{-\alpha_X}$, with $\chi^2=61.83$ for 67 degrees of freedom. The XRT fluxes were converted to monochromatic flux densities at 1~keV. Details of the XRT flux-density conversion, temporal binning, and broadband SED reconstruction are provided in Appendix~\ref{app:optical_xrt}.

To test for a late-time steepening, we used the native,
absorption-corrected 0.3--10~keV XRT fluxes rather
than the reconstructed 1-keV light curve.
Their count-to-flux conversion tracks the observed
hardness ratio.  Excluding exposures flagged by the UKSSDC for unreliable centroiding, a single power law gives $\alpha_{X,\mathrm{band}}=1.55^{+0.04}_{-0.04}$ and $\chi^2/\mathrm{dof}=66.32/67$. A continuous broken power law constrained to steepen improves the statistic by only $\Delta\chi^2=0.17$ for two additional parameters. Including flagged exposures increases the improvement to 5.41, but the break time remains poorly constrained. These products therefore do not establish a robust late-time X-ray break. The selection, fitting procedure, and sensitivity checks are described in Appendix~\ref{app:xrt_late_break_test}.

The lower panel of Figure~\ref{fig:lc_sed} shows the optical-to-X-ray SEDs reconstructed for $0.18<t<0.20$, $0.85<t<1.00$, and
$2.30<t<2.70$~d after the GBM trigger. These intervals are marked by shaded regions in the upper panel. The optical/NIR measurements were shifted to a common representative epoch within each interval,
at which the XRT flux density was evaluated from
the late-time temporal fit (Appendix~\ref{app:optical_xrt}). Photometric upper limits are shown for completeness but were not included in the spectral fits.

Table~\ref{tab:optical_xrt_sed} summarizes the reconstructed SEDs.
The optical/NIR-only spectral indices are flatter than those obtained when XRT is included. The joint single-power-law fits yield
$\chi^2/{\rm dof}=17.41/1$, $8.80/1$, and $4.70/2$ for the three epochs, respectively, providing a poor description at the first two epochs under the adopted corrections and uncertainties. The XRT flux densities are predictions from the same late-time temporal fit and are therefore correlated; the latter two epochs lie within XRT coverage gaps.

\section{Time-resolved gamma-ray spectra and LAT high-energy events}
\label{sec:spectra}

We examine the gamma-ray spectra across the main
emission episodes and the intervening activity,
together with the energies and arrival times of
LAT photons. These results complement the temporal
analysis in Section~\ref{sec:temporal} and provide spectral constraints for the modelling in Section~\ref{sec:description}.

\subsection{BAT spectra across the early episodes}
\label{sec:bat_spectra}

Table~\ref{tab:bat_pl} summarizes the power-law fits to the
15--150~keV BAT spectra over 70--165, 250--320,
and 334--390~s after the GBM trigger. The inter-episode
emission has a lower energy flux and a softer
best-fitting spectrum than either bright episode,
although its photon index is less precisely constrained.

For all three intervals, the cutoff-power-law
fits converge to the power-law limit.
The BAT spectra therefore do not constrain
a finite cutoff energy or provide a BAT-only
measurement of $E_{\rm peak}$.

We also tested for spectral evolution within the two bright
BAT intervals using shorter spectra and their corresponding
responses (Fig.~\ref{fig:BAT_photon}). In the first episode,
the fitted power-law index
increased from $1.07^{+0.03}_{-0.03}$ at 70-90\,s to
$1.21^{+0.04}_{-0.04}$ at 140-165\,s. During the second
pulse, it increased from $1.00^{+0.05}_{-0.05}$ at
334-344\,s to $1.50^{+0.30}_{-0.29}$ at 379-390\,s.
Uncertainties are quoted at 90 per cent confidence.
Compared with a shared photon index within each
episode, allowing independent indices in its
subintervals improves the fit by $\Delta\chi^2=31.93$
for three additional parameters in the first episode
and $\Delta\chi^2=47.61$ for four in the second.
Each subinterval retains an independent normalization
in both comparisons. These results are conditional
on the adopted diagonal treatment of systematic
uncertainties.

These results support spectral softening within
both bright episodes, whereas Table~\ref{tab:bat_pl} reports
their time-averaged spectral properties.

\subsection{LAT high-energy events and spectral evolution}
\label{sec:lat_photons}

The bottom-left panel of Figure~\ref{fig:LAT_lightcurve} shows the energies and arrival times of the selected LAT events.
Among those with $P_{\rm GRB}\geq0.9$, 28 exceed
1~GeV and four exceed 10~GeV. The earliest event satisfying $P_{\rm GRB}\geq0.9$ arrived at 70.39~s with an energy of 0.306~GeV. The highest-energy event, 43.31~GeV, arrived at
82.24~s. Other prominent events include 24.95~GeV
at 122.92~s and 14.24~GeV at 304.15~s, the latter
between the two bright gamma-ray episodes.
Table~\ref{tab:lat_high_energy_events}  lists the ten highest-energy events and
their source-association probabilities.
Their physical interpretation is discussed in
Section~\ref{sec:maximum_synchrotron_energy}.

The middle-left panel of Figure~\ref{fig:LAT_lightcurve} shows the fitted power-law spectra for the five principal intervals,
expressed as $\nu F_\nu\propto E^{2-\Gamma_{\rm LAT}}$
over 0.1--1000~GeV. The corresponding photon indices
are shown in the middle-right panel and listed in
Table~\ref{tab:lat_integrated_spectra}. The model envelopes reflect the local normalization--index covariance.

A joint likelihood fit with a shared photon index
across the five intervals gives
$\Gamma_{\rm LAT}=2.24\pm0.06$, with nominal
one-parameter profile uncertainties defined by
$2\Delta\ln\mathcal{L}=1$.
Allowing independent indices improves the fit by
only $2\Delta\ln\mathcal{L}=0.63$ for four additional
parameters. The spectra are therefore compatible
with a shared photon index at the adopted temporal
resolution. Sensitivity tests are described in
Appendix~\ref{app:lat_systematics}.

%We compared the sum of the five original LAT log likelihoods under a common photon index with the corresponding fit allowing five independent indices.  The fit with a shared photon index gives$\Gamma_{\rm LAT}=2.24^{+0.06}_{-0.06}$, where the errors are nominal profile intervals defined by $2\Delta\ln L=1$. Allowing four additional index parameters improves the fit by only $2\Delta\ln L=0.63$, corresponding to a nominal asymptotic $\chi^2_4$ tail probability of 0.96. The five interval-resolved spectra are therefore compatible with a common photon index at the adopted temporal resolution. The result is stable between diffuse-normalization ceilings of 30 and 100. The low-count sampling distribution was not calibrated by simulation.

Over the same intervals, the 0.1--1000~GeV energy
flux decreases from $(47.43\pm8.95)\times10^{-8}$
to $(0.93\pm0.52)\times10^{-8}$ erg~cm$^{-2}$~s$^{-1}$ (Table~\ref{tab:lat_integrated_spectra}).

\subsection{GBM spectra of the bright episodes}

The left panel of Figure~2 shows the GBM spectrum
integrated over the first bright episode,
from $-6.1$ to $164.9$~s relative to the GBM trigger. With the adopted linear background, adding a blackbody
to the Band model improves the fit by
$\Delta{\rm PG}=20.94$ for two additional parameters,
with $k_{\rm B}T=42.7^{+3.5}_{-3.4}$~keV (Table~\ref{tab:gbm_spectral_parameters}). The sensitivity to the background prescription
is discussed in Appendix~\ref{app:gbm_systematics}.

For the early $-6.1$--50~s interval, adding a BB
improves the fit by $\Delta{\rm PG}=16.93$ relative
to either Band or Comptonized with the adopted linear
background, with $k_{\rm B}T\simeq33$~keV for Band+BB.
Here, $\Delta{\rm PG}$ denotes the decrease in PG
after adding the BB component. The Band+BB high-energy
index reaches its lower bound, effectively approaching
a Comptonized+BB spectrum over the observed energy
range. Tables~\ref{tab:gbm_spectral_parameters} and~\ref{tab:gbm_spectral_normalizations} report the spectral parameters
and normalizations, and Figure~\ref{fig:gbm_spectra_eps1} shows the fits.

Splitting this interval at 12.29~s also yields fit
improvements, but competing temperature solutions
prevent a unique determination of the BB temperature
evolution. In the subsequent 50--115 and
115--164.9~s intervals, adding BB to Band gives
$\Delta{\rm PG}=6.88$ and 1.11, respectively,
with a linear background. We display the Band fit
for the final interval, where the BB parameters
are poorly constrained.

The fit improvements depend on the background prescription and the allowed
BB normalization range. We therefore do not assign
a calibrated detection significance to the additional
component. These tests are described in Appendix~\ref{app:gbm_systematics}.

\begin{figure*}
\centering
\includegraphics[width=0.95\linewidth]{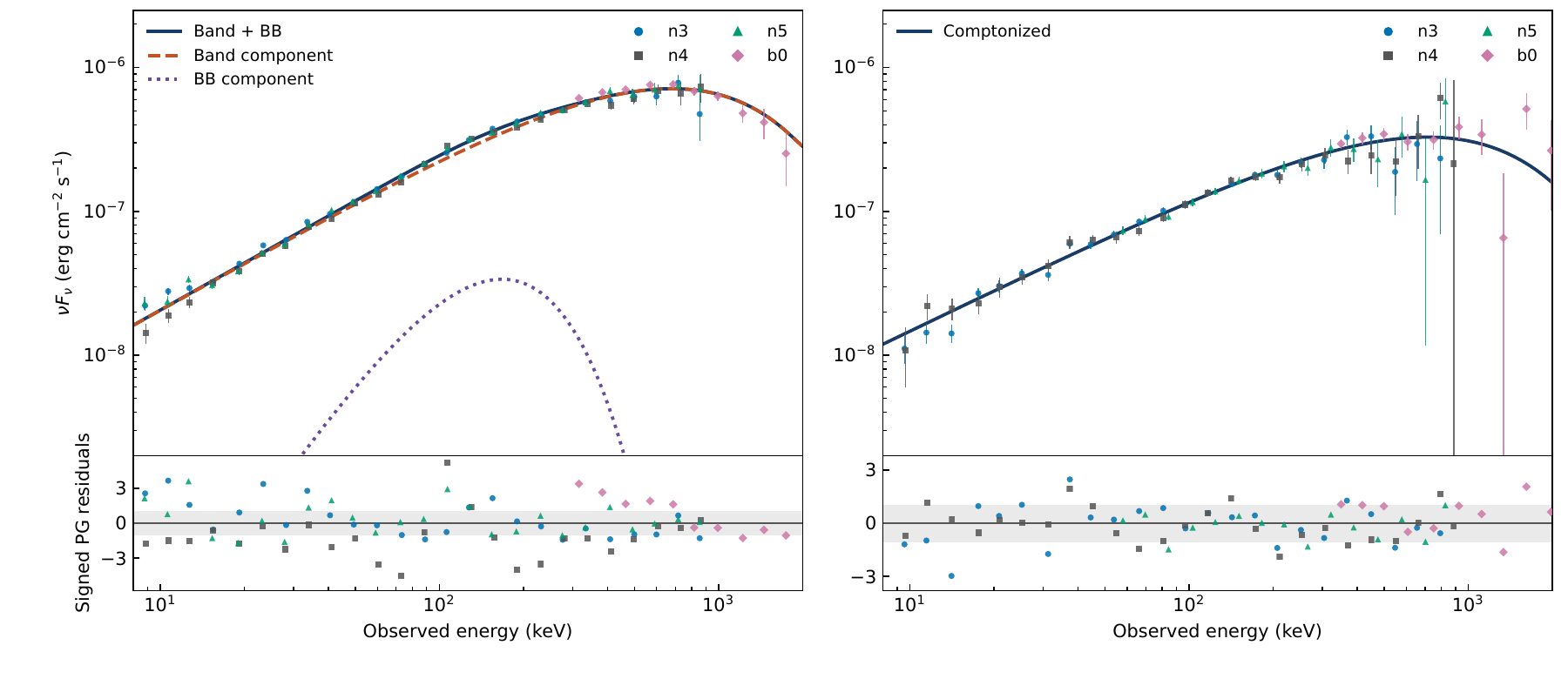}
\caption{Fermi/GBM spectra of the first
($-6.1$--$164.9$~s; left) and second
($332.8$--$390$~s; right) bright episodes,
relative to the GBM trigger.
Curves show Band+BB and its components
(left) and Comptonized (right).
Points are model-dependent unfolded $\nu F_\nu$
estimates from detectors n3, n4, n5, and b0.
Fits use count-space data, detector responses,
and linear backgrounds; the strips beneath each spectrum show signed
Poisson--Gaussian deviance residuals. The left panels share identical display grouping, while all 485 original channels enter the fit.
The display ends at 2~MeV. BB is included as an empirical spectral component.
}
\label{fig:gbm_spectra}
\end{figure*}

\begin{figure*}
\centering
\includegraphics[width=0.95\linewidth]{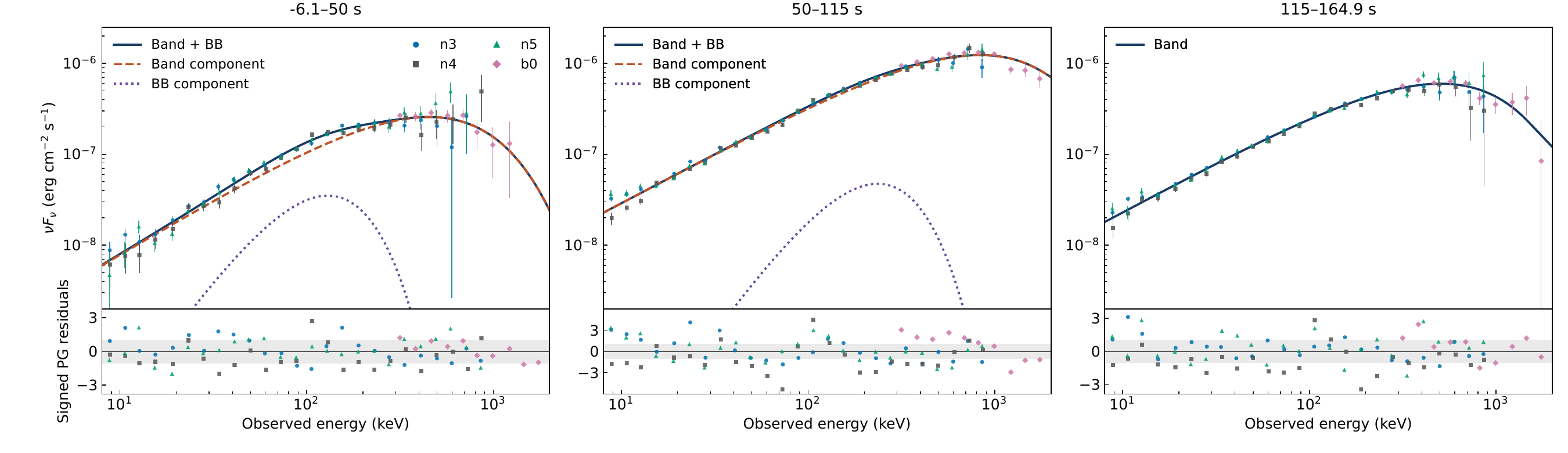}
\caption{Time-resolved \textit{Fermi}/GBM spectra during the first
bright episode of GRB~260924A. Panels cover $-6.1$--50, 50--115, and 115--164.9~s after the GBM trigger. The first two show Band+BB fits and their components; the third shows a Band fit.
The strips beneath each spectrum show signed
Poisson--Gaussian deviance residuals.
Non-positive background-subtracted estimates are
not displayed on the logarithmic spectral axes,
but their count-space residuals are retained. Detectors, unfolding, background treatment, and display range follow Figure~\ref{fig:gbm_spectra}. }
\label{fig:gbm_spectra_eps1}
\end{figure*}

For the second bright episode, the 332.8--390~s
spectrum is described by a Comptonized model
(Figure~\ref{fig:gbm_spectra}, right panel), with
$\alpha_{\rm GBM}=-1.05\pm0.03$,
$E_{\rm peak}=714^{+64}_{-55}$~keV, and
${\rm PG}=355.92$ for 391 fitted channels under
the linear-background prescription.
The parameter uncertainties are one-parameter
profile intervals at $\Delta{\rm PG}=1$.

Replacing Comptonized with Band improves the
statistic by $\Delta{\rm PG}=6.77$, but the
high-energy index is not bounded on both sides
within the tested domain, and the improvement
depends on the background and detector selection.
Band+BB fits also remain sensitive to the imposed
normalization bound and do not provide a stable
additional-component estimate. We therefore retain
Comptonized as the reference description.
The overlapping 334--390~s extraction gives consistent
parameters and serves as a sensitivity check
(Appendix~\ref{app:gbm_systematics}).

Adding a BB improves some of the early GBM fits,
but these comparisons do not uniquely identify
the origin of the additional spectral curvature.

\section{Description of the multiwavelength afterglow observations }
\label{sec:description}

We use external-shock models to interpret the early
and late afterglow phases identified in Section~\ref{sec:lc_sed}.
We first examine the late optical and X-ray emission
to guide the choice of circumburst density profile,
and then consider the early X-ray and LAT emission.
We connect these stages by modelling the X-ray
rebrightening through energy injection and evolving
shock microphysics.

\subsection{Late afterglow}
\label{sec:late_afterglow}

The late optical and X-ray observations provide temporal
and spectral constraints on the afterglow interpretation.
Using the convention $F_\nu\propto t^{-\alpha}\nu^{-\beta}$,
the separate light-curve fits in
Section~\ref{sec:lc_sed} give
$\alpha_{\rm opt}=1.46\pm0.01$ and
$\alpha_X=1.57\pm0.04$.
Complementary spectral information comes from the
optical-to-X-ray SEDs shown in the lower panel of Figure~\ref{fig:lc_sed}.
Single-power-law fits yield
$\beta_{\rm opt/NIR/X}=0.85\pm0.01$,
$0.89\pm0.01$, and $0.90\pm0.01$
at 0.18--0.20, 0.85--1.00, and 2.30--2.70~d,
respectively (Table~\ref{tab:optical_xrt_sed}).

We examine whether these temporal and spectral properties
are compatible with synchrotron emission in which the
optical and X-ray bands sample the same power-law segment.
In the slow-cooling regime, this would correspond to
either $\nu^{\rm syn}_m<\nu_{\rm opt}<\nu_X<\nu^{\rm syn}_c$ or
$\nu^{\rm syn}_m<\nu^{\rm syn}_c<\nu_{\rm opt}<\nu_X$,
where $\nu^{\rm syn}_m$ and $\nu^{\rm syn}_c$ denote the characteristic
and cooling synchrotron frequencies, respectively.

For a power-law distribution of accelerated electrons
with index $p$, the synchrotron spectral indices are
$\beta=(p-1)/2$ below the cooling frequency and
$\beta=p/2$ above it. If the broadband indices are identified with a single synchrotron segment, the former interpretation gives
$p=2\beta+1\simeq2.7$--$2.8$, whereas the latter gives
$p=2\beta\simeq1.7$--$1.8$.  For $p\simeq2.7$--$2.8$, a uniform medium predicts
$\alpha\simeq1.28$--$1.35$, whereas a steady stellar
wind predicts $\alpha\simeq1.78$--$1.85$. Within the $p>2$ prescription considered here,
this comparison motivates exploring
$\nu^{\rm syn}_m<\nu_{\rm opt}<\nu_X<\nu^{\rm syn}_c$.
This identification is treated as a working hypothesis,
given the departures from a single power law in the
first two optical-to-X-ray SEDs (Table~\ref{tab:optical_xrt_sed}).

In this spectral regime, the temporal decay provides
a diagnostic of the circumburst density profile,
parametrized as $n(R)\propto R^{-k}$.
The cases $k=0$ and $k=2$ correspond to a uniform
medium and a steady stellar wind, respectively,
while $0<k<2$ describes an intermediate density gradient.
For an adiabatic forward shock before a jet break,
the temporal and spectral indices satisfy \citep{Yi2013}

\begin{eqnarray}
\label{eq:k_p}
\alpha^{\rm syn}
=
\frac{3(p-1)}{4}+\frac{k}{2(4-k)},
\qquad
\beta=\frac{p-1}{2}\,.
\end{eqnarray}

%For the electron indices inferred above, $p\simeq2.71$--$2.79$, a uniform medium predicts $\alpha\simeq1.28$--$1.34$, whereas a steady stellar wind predicts $\alpha\simeq1.78$--$1.84$. The measured optical and X-ray decay indices lie between these predictions, motivating the exploration of an intermediate density profile.
Figure~\ref{fig:optical_xrt_closure} illustrates the dependence of the predicted temporal index on $p$ and $k$, with contours marking the representative optical and X-ray decay indices.
Within the stated assumptions, this comparison
provides a physical motivation for considering
a stratified circumburst medium.

Combining the reference temporal index
$\alpha_{\rm ref}=(\alpha_{\rm opt}+\alpha_X)/2$
with the observed broadband spectral indices, the closure relations (Equation~\ref{eq:k_p}) support an intermediate density gradient, with $k\simeq1.0$--$1.3$. We use $k=1.2$, corresponding
to $\beta\simeq0.87$ and $p\simeq2.73$, in the
modelling below. Here, $\alpha_{\rm ref}$ represents the midpoint
between the separately fitted temporal indices.

%To select a representative density profile, we use $\alpha_{\rm ref}=(\alpha_{\rm opt}+\alpha_X)/2\simeq1.51$. For $k=1.2$, Equation~(\ref{eq:k_p}) gives $p\simeq2.73$ and $\beta\simeq0.86$, within the range of broadband spectral indices reported above. We therefore fix $k=1.2$ in the modelling below. This choice is guided by the closure relations rather than a joint fit to the optical and X-ray temporal slopes. The connection between the early and late stages is modelled through energy injection and evolving microphysics in Section~\ref{sec:afterglow_transition}.

The observational constraints on a possible jet break
and their implications for the jet geometry are
discussed in Section~\ref{sec:jet_break_constraints}.

\subsection{Early afterglow}
\label{sec:early_afterglow}

The joint analysis of three XRT spectra extracted
outside the flare interval, with a shared intrinsic
absorption column, gives
$\Gamma_{\rm XRT}=1.78^{+0.07}_{-0.07}$
during 263.5--314.0~s (WT),
$2.25^{+0.15}_{-0.14}$
during 3652.5--3757.0~s (WT), and
$2.01^{+0.10}_{-0.10}$
during 3763.4--4656.0~s (PC),
with times measured from the GBM trigger
(Table~\ref{tab:xrt_common_absorption} and Figure~\ref{fig:XRT_spectra}). The earliest WT spectrum is harder than the two
subsequent spectra, whose central photon indices do not indicate monotonic softening. For forward-shock synchrotron emission at
$\nu_X>\max(\nu^{\rm syn}_m,\nu^{\rm syn}_c)$,
the expected photon index is $\Gamma_X=1+p/2$.
The standard adiabatic closure relation before
a jet break gives $\alpha_X=(3p-2)/4$, independent
of the circumburst density index $k$ in this
spectral regime.  For $p\simeq2.35$--$2.4$, these relations predict
$\Gamma_X\simeq2.18$--$2.20$ and
$\alpha_X\simeq1.26$--$1.30$.
The spectral prediction is consistent with the
returning WT measurement, while the subsequent
PC spectrum is somewhat harder. The predicted
temporal decay is close to that of the third
segment of the catalogue XRT fit
(Table~\ref{tab:xrt_catalogue_temporal}).

The LAT observations provide an additional test of
the forward-shock interpretation.
The light curve over 164.9--1100~s decays with
$\alpha_{\rm LAT}=1.30^{+0.20}_{-0.19}$
(Section~\ref{sec:lat_lightcurve}), while the five principal spectral
intervals are compatible with a shared photon index
$\Gamma_{\rm LAT}=\beta_{\rm LAT}+1=2.24^{+0.06}_{-0.06}$
(Section~\ref{sec:lat_photons}). We consider synchrotron and SSC emission above their
respective characteristic and cooling frequencies,
$\nu_{\rm LAT}>\max(\nu_m^{i},\nu_c^{i})$,
where $i={\rm syn,SSC}$.
In this spectral segment, both mechanisms predict
$\beta^{\rm i}=p/2$.
For an adiabatic, decelerating forward shock before
a jet break, the temporal indices are

\begin{eqnarray}
%\begin{aligned}
\alpha^{\rm syn} = \frac{3p-2}{4}, \hspace{1cm}
\alpha^{\rm SSC} = \frac{(18-5k)p+6k-20}{4(4-k)}.
%\end{aligned}
\end{eqnarray}
The SSC relation assumes scattering in the Thomson
regime.  For $p\simeq2.35$--$2.4$ and $k=1.2$, these
relations predict $\alpha^{\rm syn}\simeq1.26$--$1.30$,
$\alpha^{\rm SSC}\simeq1.38$--$1.43$, and
$\Gamma\simeq2.18$--$2.20$ for both mechanisms,
consistent with the measured LAT temporal and
spectral indices.

A common electron index ($p\simeq2.35$--$2.4$) can therefore
account approximately for the representative X-ray
and LAT decay slopes through forward-shock
synchrotron and SSC emission, respectively, above
their corresponding characteristic and cooling
frequencies. This interpretation does not fully
reproduce the XRT spectral evolution. The closure
relations alone also allow synchrotron emission
in the LAT band; the highest-energy photons provide
additional constraints on the emission mechanism
(Section~\ref{sec:maximum_synchrotron_energy}). The initial LAT excess is considered
separately in Section~\ref{sec:early_lat_excess}.

\subsection{The X-ray rebrightening: connecting the early and late afterglow}
\label{sec:afterglow_transition}

The closure-relation comparisons in Sections~\ref{sec:late_afterglow}
and~\ref{sec:early_afterglow} suggest representative electron indices
of $p\simeq2.73$ for the late afterglow and
$p\simeq2.35$--$2.4$ for the early afterglow.
Motivated by these comparisons and the intervening
X-ray rebrightening, we model the connection between
the two phases through energy injection and evolving
shock microphysics.

Slower ejecta can catch up with the decelerating blast wave
and transfer additional energy to the shocked flow through
refreshed shocks \citep{Sari_2000}. Evolution of the
effective electron and magnetic energy fractions could also
modify the emission during this transition. Such variations
are additional phenomenological assumptions, rather than
a necessary consequence of energy injection \citep{2014ApJ...787..168V}.
Variable-microphysics models illustrate their possible
effects on afterglow temporal and spectral behaviour
\citep[e.g., see][]{2006A&A...458....7I, 2006MNRAS.369..197F, 2006MNRAS.369.2059P, 2013MNRAS.428..845L, 2024MNRAS.527.1884F}.

The rebrightening is shown in Figure~\ref{fig:XRT_lightcurve} and described
by the fourth segment of the \textit{Swift}/XRT catalogue
fit (Table~\ref{tab:xrt_catalogue_temporal}). We adopt $t_a=4700$~s and $t_b=6200$~s
after the GBM trigger as the boundaries of the
modelled transition. We denote by $\epsilon_{e,f}$ and $\epsilon_{B,f}$
the fractions of the post-shock internal energy
carried by accelerated electrons and magnetic
fields, respectively, where the subscript $f$
refers to the forward shock. During $t_a<t<t_b$, we adopt

{\small
\begin{eqnarray}\nonumber
%\begin{aligned}
p_{\rm }(t) = p\left(\frac{t}{t_a}\right)^{s_p},\,\,\, 
\epsilon_{e_f}(t) =\varepsilon_{e_f}\left(\frac{t}{t_a}\right)^{-\alpha_e},\,\,\,
\epsilon_{B_f}(t) =\varepsilon_{B_f}\left(\frac{t}{t_a}\right)^{-\alpha_B}.
%\end{aligned}
\end{eqnarray}
}
Here, $p$, $\varepsilon_{e_f}$, and $\varepsilon_{B_f}$
denote the pre-transition electron index and
microphysical energy fractions of the early
afterglow model. The indices $\alpha_e$ and $\alpha_B$ describe
the temporal evolution of the electron and 
magnetic energy fractions, respectively. The exponent $s_p=\ln(p_{\rm late}/p)/\ln(t_b/t_a)$
connects $p$ to the adopted late-time value
$p_{\rm late}=2.75$, within the range inferred
in Section~\ref{sec:late_afterglow}.  For $t\leq t_a$, these quantities retain their
pre-transition values; for $t\geq t_b$, they remain
fixed at the values reached at $t_b$.

We denote by $E_{\rm K,iso}$ the isotropic-equivalent
kinetic energy before the rebrightening and by
$E_{\rm K,iso}(t)$ its subsequent evolution.
Energy injection is restricted to $t_a<t<t_b$,
such that
\begin{eqnarray}\nonumber
\frac{E_{\rm K,iso}(t)}{E_{\rm K,iso}} =
\begin{cases}
1, & t\leq t_a,\\[2pt]
(t/t_a)^{q_{\rm inj}}, & t_a<t<t_b,\\[2pt]
(t_b/t_a)^{q_{\rm inj}}, & t\geq t_b.
\end{cases}
\end{eqnarray}
We adopt an energy increase by a factor
$f_E\equiv E_{\rm K,iso}(t_b)/E_{\rm K,iso}=2.52$,
which gives
$q_{\rm inj}=\ln f_E/\ln(t_b/t_a)\simeq3.34$.
This increase acts together with the prescribed
microphysical evolution to connect the early
and late afterglow flux levels.

\subsection{A reverse-shock SSC interpretation of the early LAT excess}
\label{sec:early_lat_excess}

The interval-averaged 0.1--1000~GeV photon flux
during 70.04--164.9~s exceeds the backward
extrapolation of the power-law fit to the four
subsequent LAT intervals by a factor of
$2.25\pm0.59$ (Section~\ref{sec:lat_lightcurve}).
To characterize this early excess and its steeper
decline, we fitted a continuous broken power law
with the break fixed at $t_{\rm br}=164.9$~s,
coinciding with the adopted end of the first
bright GBM/BAT episode marked in Figure~\ref{fig:gbm_lightcurve}.
The fit gives $\alpha_1=2.95^{+0.36}_{-0.35}$
and $\alpha_2=1.30^{+0.20}_{-0.19}$.  Uncertainties correspond to 68.3 per cent profile intervals obtained with $\Delta\chi^2=1$, reoptimizing
the normalization and the other temporal index
while keeping $t_b=164.9$~s fixed. The spectral models, exposure weights, and flux uncertainties were held fixed. The early index is inferred from a single
interval-averaged flux under the imposed
continuity and fixed break time.

This inferred early decline is steeper than
the forward-shock synchrotron and SSC predictions
considered in Section~\ref{sec:early_afterglow}, motivating an additional short-lived component.
We describe the early LAT excess with reverse-shock
SSC emission in the thick-shell regime, following
the stratified-medium prescriptions of
\citet{2024MNRAS.527.1674F}.
We adopt a fiducial reverse-shock crossing time
of $t_\times=150$~s after the GBM trigger.
This model timescale is distinct from the fixed
break time used to characterize the LAT light curve.

After the reverse shock crosses the ejecta,
fresh electron injection ceases and the previously
accelerated electrons continue to cool.
As the SSC cutoff moves below the LAT band,
the observed emission can become dominated by
photons arriving from larger angles to the line
of sight \citep{2000ApJ...543...66P, 2003ApJ...597..455K}. For a power-law spectral segment, the curvature
relation gives $\alpha=2+\beta$, with the temporal
index evaluated using the appropriate time origin
for the emitting episode.

The early LAT photon index,
$\Gamma_{\rm LAT}=2.21\pm0.08$, corresponds to
$\beta_{\rm LAT}=1.21\pm0.08$ and hence
$2+\beta_{\rm LAT}=3.21\pm0.08$, consistent
numerically with the inferred early decay index. This numerical agreement does not constitute a test of the curvature relation, because the post-crossing decay is not separately resolved. The combined forward- and reverse-shock model
is examined in Section~\ref{sec:cons_phys_parameter}.

\subsection{Constraints on the physical parameters}
\label{sec:cons_phys_parameter}

We use Markov chain Monte Carlo (MCMC) sampling to
explore the physical parameters of the proposed
external-shock interpretation. We first fit a
forward-shock model to the selected LAT, X-ray, and
optical/IR observations, excluding the first LAT
interval to establish a reference solution for the
longer-lived emission. The sampled parameters are
$E_{\rm K,iso}$, $n_{\rm dec}$, $p$, $\log_{10}\Gamma_0$,
$\varepsilon_{B_f}$, $\varepsilon_{e_f}$, $\zeta_{e_f}$,
$\alpha_e$, and $\alpha_B$. Here, $\Gamma_0$ is the
initial bulk Lorentz factor, and $\zeta_{e_f}$ is
the fraction of forward-shock electrons injected
into the modelled non-thermal distribution.
The parameters $\alpha_e$ and $\alpha_B$ describe
the prescribed temporal evolution of the electron
and magnetic energy fractions.

We then fit a combined forward- and reverse-shock
model, now including the first LAT interval,
with reverse-shock SSC accounting for the early
excess discussed in Section~\ref{sec:early_lat_excess}.
This model introduces $\varepsilon_{B_r}$,
$\varepsilon_{e_r}$, and $\zeta_{e_r}$ as additional
parameters, where the subscript $r$ denotes
the reverse shock.
The shared dynamical and forward-shock parameters
are sampled jointly with the reverse-shock
parameters.

We fix the redshift to $z=3.51$ and calculate the
luminosity distance for a spatially flat
$\Lambda$CDM cosmology with
$H_0=67.4~{\rm km\,s^{-1}\,Mpc^{-1}}$
and $\Omega_{\rm M}=0.315$.

The circumburst number-density profile is parametrized as $n(R)=n_{\rm dec} \left(R/R_{\rm dec}\right)^{-k}$, where $n_{\rm dec}$ is the ambient number density
at the deceleration radius. We fix $k=1.2$ and
sample $n_{\rm dec}$ directly.  Using the approximate
deceleration criterion
$\Gamma_0^2 M_{\rm sw}(R_{\rm dec})c^2
\simeq E_{\rm K,iso}$, we calculate

\begin{eqnarray}
\label{eq:r_dec}
R_{\rm dec}\simeq
\left[
\frac{(3-k)E_{\rm K,iso}}
{4\pi m_p c^2 n_{\rm dec}\Gamma_0^2}
\right]^{1/3}\,.
\end{eqnarray}

Here, $m_p$ is the proton mass and $c$ is the
speed of light. This expression follows from
$M_{\rm sw}(R_{\rm dec})=
4\pi m_p n_{\rm dec}R_{\rm dec}^{3}/(3-k)$.
The equivalent normalization in
$n(R)=AR^{-k}$ is $A=n_{\rm dec}R_{\rm dec}^{k}$.

We explored the model parameters using
\texttt{emcee} version 3.1.6
\citep{2013PASP..125..306F}.
The likelihood combined the \textit{Swift}/XRT
0.3--10~keV, \textit{Fermi}/LAT, and optical observations.
Asymmetric measurement uncertainties were
symmetrized by adopting the larger of the upper
and lower errors for each data point.
We adopted uniform priors over the ranges listed
in Table~\ref{tab5:fit}, defined in $\log_{10}\Gamma_0$
for the initial bulk Lorentz factor and in
linear space for the remaining free parameters.
The walkers were initialized around the best-fitting
parameter values obtained from a preliminary
$\chi^2$ minimization.

For the forward-shock fit, we used 128 walkers
and 10,000 steps per walker, discarding the
first 1,000 steps as burn-in.
The combined forward- and reverse-shock fit
used the same sampling configuration, with
the predicted flux given by the sum of the
two emission components.
Convergence was assessed by requiring the
post-burn-in chain length to exceed 50 times
the estimated integrated autocorrelation time
for each sampled parameter.
The mean acceptance fraction was approximately 0.3.
The parameter traces were also inspected visually
to check for stable sampling without persistent
drifts.

Table~\ref{tab5:fit} reports the posterior medians and
16th--84th percentile intervals for both fits.
Figure~\ref{tab5:fit} shows the posterior distributions
for the forward-shock-only model.
For brevity, we omit the corner plot of the
combined fit, whose shared parameter estimates
remain consistent with those of the
forward-shock-only solution.  Some posterior distributions approach the adopted prior boundaries; their quoted intervals are
conditional on these bounds.

The kinetic energy and electron spectral index
reported in Table~\ref{tab5:fit} correspond to the
pre-transition quantities $E_{\rm K,iso}$ and $p$.
The tabulated forward-shock energy fractions
correspond to $\varepsilon_{e_f}$ and
$\varepsilon_{B_f}$ in Section~\ref{sec:afterglow_transition}.
The sampled indices $\alpha_e$ and $\alpha_B$
govern their evolution between
$t_a=4700$~s and $t_b=6200$~s after the GBM trigger.
We fix $p_{\rm late}=2.75$ and $f_E=2.52$.
After $t_b$, the kinetic energy and microphysical
parameters remain constant.

Figure~\ref{fig:multi_afterglow} compares the model light curves with
the observations. The inset in the upper panel
shows the unsmoothed XRT light curve, highlighting
the rebrightening described through energy
injection and evolving shock microphysics
in Section~\ref{sec:afterglow_transition}.
The forward-shock-only solution underpredicts
the first LAT interval, whereas the combined
model accounts for this early excess through
reverse-shock SSC emission.
The inferred reverse-shock parameters are
conditional on this emission mechanism and
the adopted dynamical prescription.

\section{Physical implications of the proposed interpretation}
\label{sec:discussion}

We discuss the implications of the multiwavelength
observations and external-shock modelling for
particle acceleration, ejecta magnetization,
shock microphysics, and the circumburst environment.
We also examine the prompt radiative efficiency
and constraints on jet geometry, and compare
the gamma-ray, X-ray, and optical emission
with that of other energetic GRBs.

\subsection{Maximum synchrotron energy and 
SSC emission}
\label{sec:maximum_synchrotron_energy}

Figure~\ref{fig:synchrotron_limits} compares the energy--time distribution of the LAT events with the forward-shock maximum
synchrotron energy and an acceleration-based
reference for the reverse shock. The grey scale indicates the probability of association with GRB~260924A under the adopted
LAT source and background model.

For the acceleration-based estimates, we adopt
$t'_{\rm acc}=\eta_{\rm acc}2\pi r'_L/c$ with
$\eta_{\rm acc}=1$, corresponding to one gyration
period. Here, $r'_L$ is the electron Larmor radius.   Unless otherwise stated, photon energies and times are expressed in the observer frame;
primed quantities refer to the comoving frame of the emitting fluid. Equating this timescale to the pitch-angle-averaged
synchrotron cooling time gives
$\gamma_{\max,i}=[3e/(\eta_{\rm acc}\sigma_T B'_i)]^{1/2}$,
where $i=f,r$ denotes the forward or reverse shock,
$e$ is the elementary charge, $\sigma_T$ is the
Thomson cross-section, and $B'_i$ is the comoving
magnetic-field strength.

Using the photon-energy convention
$E'^{\rm syn}_i=\hbar eB'_i\gamma^2/(m_ec)$
and approximating the Doppler factor as
$\delta_i\simeq\Gamma_i$, we obtain
$E^{\rm syn}_{\rm max,i}(t)
\simeq 5.57\,{\rm MeV}\,\Gamma_i(t)$
for $z=3.51$ and $\eta_{\rm acc}=1$.
Here, $\Gamma_i$ is the bulk Lorentz factor of
the corresponding emitting shocked fluid,
$\hbar$ is the reduced Planck constant, and
$m_e$ is the electron mass.

For an adiabatic, constant-energy forward shock
decelerating in a medium with $n\propto R^{-k}$,
this maximum energy evolves as
$E_{\rm max,f}^{\rm syn}
\propto t^{-(3-k)/[2(4-k)]}$.  For the adopted density index $k=1.2$, this gives $E_{\mathrm{max},f}^{\mathrm{syn}}\propto
t^{-9/28}$; the wind case, $k=2$, recovers
the $t^{-1/4}$ scaling. This relation applies
during the decelerating, constant-energy phase.
Before deceleration, an approximately constant
bulk Lorentz factor gives an approximately
constant acceleration-based maximum energy
under the same assumptions.
The LAT interval considered here precedes the
prescribed energy-injection phase in Section~\ref{sec:afterglow_transition}.
The reverse-shock evolution must instead follow
the dynamics and electron population of the
shocked ejecta.

For the reverse shock, fresh electron injection
ceases at the adopted crossing time
$t_\times=150$~s (Section~\ref{sec:early_lat_excess}).
Beyond this time, the blue curve serves only
as an acceleration-based reference: it does not
represent the evolving synchrotron cutoff of
the cooling electron population or its delayed
high-latitude emission.

Several high-probability LAT events exceed the
synchrotron energies represented in Figure~\ref{fig:synchrotron_limits},
including the 43.31, 24.95, and 14.24~GeV photons
detected at 82.24, 122.92, and 304.15~s, respectively.
Together with the light-curve modelling in
Section~5.5, this supports an SSC contribution
from the forward and reverse shocks under
the adopted prescriptions \citep{2001ApJ...548..787S,2019ApJ...883..162F}.

The 14.24~GeV photon is particularly informative
because it arrived between the two main GBM/BAT
episodes. Forward-shock SSC could contribute at
this epoch, while the reverse-shock contribution
depends on its evolving electron distribution
and SSC cutoff. The arrival time alone does not
separate these possibilities or exclude a prompt
contribution, since weaker activity persists
between the bright episodes.

At $z=3.51$, attenuation by the extragalactic background
light (EBL) is relevant to the interpretation of the
highest-energy LAT photons. To estimate this propagation
effect, we use the EBL model of \citet{SaldanaLopez2021}
and its published optical-depth tables
\citep{Dominguez2024EBL}. The observed differential photon
flux is related to the intrinsic flux by
$\phi_{\rm obs}(E,t)=\phi_{\rm int}(E,t)
\exp[-\tau_{\rm EBL}(E,z)]$, where $E$ is the
observer-frame photon energy. Using the reference
cosmology of these tables, the transmissions at
14.24, 24.95, and 43.31~GeV are approximately
0.89, 0.61, and 0.27, respectively.
These transmissions characterize propagation
losses and do not constitute an EBL-corrected
refit of the broadband model.
EBL absorption reduces the transmitted flux
without changing the energies of the surviving
photons, and therefore does not alter their
comparison with the maximum synchrotron energy.

\subsubsection{Klein--Nishina effects}
\label{sec:KN}

The emission model includes Klein--Nishina effects
for both the forward and reverse shocks. We evaluate the characteristic Klein--Nishina
energies for both shock components near the
arrival times of the highest-energy LAT photons.

For the forward shock, the observer-frame
values are $E^{\rm SSC}_{\rm KN,c, f}\simeq86$~GeV and
$E^{\rm SSC}_{\rm KN,m, f}\simeq252$~GeV at 80~s,
and approximately 99 and 136~GeV at 300~s,
respectively.
The 43.31~GeV and 14.24~GeV photons therefore lie
below both characteristic scales evaluated near
their respective arrival times.
These estimates do not indicate a strong
Klein--Nishina suppression at those photon energies
within the adopted forward-shock prescription.

For the reverse shock, the observer-frame Klein--Nishina characteristic energies at $t\simeq80$~s are $E^{\rm SSC}_{\rm KN,c, r}\simeq38$~GeV and $E^{\rm SSC}_{\rm KN,m, r}\simeq150$~GeV.
At the adopted crossing time, $t_\times=150$~s, these values are approximately 40 and 150~GeV, respectively.  The 43.31~GeV photon detected at 82~s lies slightly above the lower of these scales.
Klein--Nishina suppression may therefore affect its interpretation as reverse-shock SSC emission, although these characteristic energies do not represent sharp spectral cutoffs.

\subsection{Ejecta magnetization }
\label{sec:photosphere_magnetization}

The reverse-shock SSC interpretation of the early
LAT excess provides a way to examine the magnetic
properties of the shocked ejecta.
A magnetic field carried by the outflow could
contribute to the synchrotron emission that supplies
the seed photons for this SSC component.

For the combined forward- and reverse-shock fit
(Table~\ref{tab5:fit}), the posterior median magnetic-energy
fractions are $\varepsilon_{B_r}=9.78\times10^{-5}$
and $\varepsilon_{B_f}=1.42\times10^{-5}$.
Their ratio is approximately 6.9, giving

\begin{eqnarray}
R_B\equiv
\left(\frac{\varepsilon_{B_r}}
{\varepsilon_{B_f}}\right)^{1/2}
\simeq2.6.
\end{eqnarray}

This value is calculated from the reported
posterior medians.
Although the two components are constrained
mainly at different epochs, the constant
pre-transition forward-shock fraction adopted
in Section~\ref{sec:afterglow_transition} allows this comparison near
shock crossing.
In the hydrodynamic limit, where the post-shock
internal-energy densities are comparable,
$B'_r/B'_f\simeq R_B$.

The larger reverse-shock magnetic-energy fraction
is compatible with a magnetic field retained
by the ejecta \citep{2003ApJ...595..950Z}, although different efficiencies
of magnetic-field amplification at the two
shocks could also contribute to the contrast.
This inference depends on the reverse-shock
SSC interpretation of the early LAT excess.

For cold, unshocked ejecta, the magnetization
is defined as
$\sigma_{\rm ej}=B_{\rm ej}^{\prime\,2}/
(4\pi\rho'_{\rm ej}c^2)$,
where $B'_{\rm ej}$ and $\rho'_{\rm ej}$ are
the comoving magnetic field and rest-mass density.
This upstream quantity is distinct from
$\varepsilon_{B_r}$ and $R_B$.
Connecting them requires a shock model that
accounts for compression, magnetic stresses,
and energy dissipation \citep{2005ApJ...628..315Z}.
The inferred contrast therefore supports
magnetized ejecta within the adopted interpretation,
but does not measure $\sigma_{\rm ej}$ or establish
Poynting-flux dominance.

\subsection{Electron energy partition and
accelerated-electron fractions}
\label{sec:frac_ele}

For the combined forward- and reverse-shock fit
in Table~\ref{tab5:fit}, the inferred electron energy fractions
are $\varepsilon_{e_r}=0.204^{+0.021}_{-0.020}$ and
$\varepsilon_{e_f}=0.0211^{+0.0019}_{-0.0018}$.
The ratio of these medians is
$\mathcal{R}_e\equiv
\varepsilon_{e_r}/\varepsilon_{e_f}\simeq9.7$.
Within the adopted model, a larger fraction of
the local post-shock internal energy is therefore
assigned to non-thermal electrons in the shocked
ejecta. 

A useful comparison is provided by
\citet{2024MNRAS.527.1674F}, who modelled eight
bright Fermi-LAT GRBs and obtained
$2.8\times10^{-3}\lesssim\varepsilon_{e_f}\lesssim0.12$
and $0.03\lesssim\varepsilon_{e_r}\lesssim0.13$.
Our forward-shock value lies within their reported
range, while the reverse-shock value is approximately
1.6 times their largest reported value.
These ranges provide modelling context rather
than universal bounds.
Earlier modelling of GRB~990123 also allowed
substantially different microphysical parameters
behind the two shocks
\citep{2004MNRAS.353..511P}.

Using a sample of 49 radio afterglows,
\citet{2023MNRAS.518.1522D} combined radio-peak
diagnostics with prompt-efficiency considerations
to obtain approximate ranges
$0.01\lesssim\varepsilon_e\lesssim0.2$ and
$0.1\lesssim\zeta_e\lesssim1$.
Our forward-shock values fall within these ranges.
This comparison applies to the shocked ambient
medium and does not impose the same bounds on
the reverse-shock population.

The electron energy fractions should be interpreted
together with the accelerated-electron fractions,
$\zeta_{e_r}=0.0342$ and $\zeta_{e_f}=0.89$.
Within the adopted prescription, approximately
3.4 per cent of the reverse-shock electrons carry
20.4 per cent of the local post-shock internal
energy at injection.
The corresponding ratios,
$\varepsilon_{e_r}/\zeta_{e_r}\simeq5.96$ and
$\varepsilon_{e_f}/\zeta_{e_f}\simeq2.37\times10^{-2}$,
describe how this energy is distributed among
the accelerated electrons.
At fixed dissipated energy per particle and
electron spectral index, a smaller accelerated
fraction increases the characteristic injection
energy.
These ratios do not directly determine
$\gamma_{m_r}/\gamma_{m_f}$, since the two shocks
have different strengths and injection conditions.
Electrons outside the modelled power-law
distribution need not be cold or radiatively
inactive.

A relevant precedent is \citet{2023MNRAS.525.5224M}, who
adopted accelerated-electron fractions of 0.01
in the reverse shock and unity in the forward
shock for GRB~030329, rather than fitting them
independently.
Allowing incomplete electron injection introduces
degeneracies among the energy, density, and
microphysical parameters in standard afterglow
models \citep{2005ApJ...627..861E}.
Our inferred fractions are therefore conditional
on the adopted priors and emission prescriptions,
including the reverse-shock SSC origin assigned
to the early LAT excess.

\subsection{Density profile of the circumburst medium}

The temporal and spectral constraints discussed
in Section~\ref{sec:late_afterglow} favour an intermediate density
gradient, $k\simeq1.0$--$1.3$.
We use $k=1.2$ within this range and hold it
fixed during the parameter fitting.
The combined forward- and reverse-shock fit
gives $n_{\rm dec}=2.51^{+0.51}_{-0.46}$~cm$^{-3}$
(Table~\ref{tab5:fit}), evaluated at the deceleration radius
calculated for each sampled parameter set.

This stratification is intermediate between a
homogeneous medium ($k=0$) and a steady stellar
wind ($k=2$). Comparable profiles have been
inferred for other GRBs.
\citet{Yi2013} modelled the early
optical afterglows of 19 GRBs and obtained
$0.4\leq k\leq1.4$, with a typical value near unity.
Similarly, \citet{2013ApJ...774...13L} identified
38 onset components in a parent sample of
146 optical afterglows and inferred a
characteristic profile $n\propto R^{-1}$.
For GRB~130427A, \citet{2013ApJ...779L...1K}
obtained $k=1.4\pm0.2$ from the temporal and
spectral behaviour within their synchrotron
interpretation.

Physically, a profile shallower than $R^{-2}$
could reflect non-steady mass loss or modifications
of the circumstellar wind structure.
Interactions with the surrounding medium and
the formation of a shocked-wind bubble provide
possible ways to alter the density profile
\citep{2006A&A...460..105V}.    We therefore interpret $k=1.2$ as an effective description of the external density over the
radii probed by the observations, without
identifying a unique mass-loss history or
a wind-termination shock.
This external profile governs the coupled
forward- and reverse-shock dynamics; it does
not describe the density distribution within
the ejecta.

\subsection{Prompt energetics and radiative efficiency}
\label{radiat_eff}

We estimate the prompt radiative efficiency using
\begin{equation}
\eta_\gamma =
\frac{E_{\gamma,\mathrm{iso}}}
     {E_{\gamma,\mathrm{iso}}+E_{\rm K, iso}}\,.
\end{equation}

The published prompt energy,
$E_{\gamma,\mathrm{iso}}\simeq9.74\times10^{54}$~erg
in the rest-frame 1--10,000~keV band
\citep{GCN45750}, becomes
$9.79\times10^{54}$~erg when rescaled to our adopted
cosmology. An independent analysis of the GBM CSPEC data with
RMfit, using non-overlapping time intervals, gives
$E_{\gamma,\mathrm{iso}}\simeq9.31\times10^{54}$~erg
for Band fits and $9.37\times10^{54}$~erg for
Comptonized fits in the same rest-frame band.
These estimates are approximately 4--5 per cent
below the adopted value and provide an independent
check of the burst's extreme energetics. Combining the adopted value of
$E_{\gamma,\mathrm{iso}}=9.79\times10^{54}$~erg
with $E_{\rm K, iso}=7.01^{+0.39}_{-0.37}\times10^{55}$~erg from
the joint forward- and reverse-shock fit
(Table~\ref{tab5:fit}), we obtain
$\eta_\gamma=0.123^{+0.006}_{-0.006}$. The quoted uncertainty propagates only the
reported uncertainty in $E_{\rm K,iso}$,
holding $E_{\gamma,\rm iso}$ fixed, and
excludes model systematics.

This efficiency refers to the prompt emission
and the pre-injection kinetic energy; it excludes
the energy subsequently supplied through the
prescription in Section~\ref{sec:afterglow_transition}.
The initial kinetic energy is approximately
7.2 times the prompt gamma-ray energy.
Our inferred efficiency lies within the range
$0.01\lesssim\eta_\gamma\lesssim0.36$ reported
by \citet{2024MNRAS.527.1674F} for eight \textit{Fermi}/LAT bursts.
Within the adopted model, the extreme prompt
energy is therefore accompanied by a large
kinetic-energy reservoir rather than requiring
an unusually high radiative efficiency.

An efficiency of this order is compatible with internal-shock calculations that allow a substantial dispersion in the Lorentz factors of the colliding shells \citep{2001ApJ...557..399G}, although it does not uniquely identify the prompt dissipation mechanism.
Its interpretation also depends on the adopted
afterglow model and the treatment of radiative
losses \citep{2007ApJ...655..989Z}. The forward and reverse shocks share the dynamical energy budget of the ejecta, so
their kinetic energies are not added as
independent reservoirs in this estimate.

The inferred prompt energy places GRB~260924A
among the most energetic GRBs known, comparable
in isotropic-equivalent energy to GRB~221009A,
for which estimates of approximately
$(1.0$--$1.2)\times10^{55}$~erg have been reported \citep{Lesage2023,Frederiks2023}.
Other exceptionally energetic events include
GRB~080916C, GRB~090323, GRB~090902B, and
GRB~220101A \citep{Beniamini2015,Scotton2023}.
These comparisons place GRB~260924A at the extreme
high-energy end of the observed population, although
differences in spectral integration bands and temporal
coverage preclude a precise ranking from the published
values alone.

The afterglow modelling likewise implies a
substantial isotropic-equivalent kinetic-energy
reservoir. Estimates of several $10^{55}$~erg have
previously been obtained from the GeV afterglows of
GRB~090323 and GRB~090902B under a forward-shock
synchrotron interpretation \citep{Beniamini2015}.
The quantitative comparison depends on the adopted
radiation mechanism and microphysical parameters.
Furthermore, the absence of a robust jet-break
measurement in our analysis leaves the collimation
correction uncertain. The extreme isotropic-equivalent
energetics therefore do not, by themselves, establish
an equally exceptional beaming-corrected energy.

\subsection{Constraints on a possible jet break}
\label{sec:jet_break_constraints}

The late-time XRT tests described in Section~\ref{sec:lc_sed}
and Appendix~\ref{app:xrt_late_break_test} do not establish a robust
steepening. The improvement obtained by introducing
a break depends on the data selection, and its
epoch remains poorly constrained.
The optical observations provide no evidence
for a contemporaneous, sustained steepening
that would support an achromatic jet break.

A jet break remains possible, but the available
observations provide neither a measured break
time nor a robust lower bound on it.
Consequently, we do not infer a jet opening angle
or a beaming-corrected energy from these data.
The true energy budget therefore remains uncertain
despite the extreme isotropic-equivalent
energetics discussed in Section~\ref{radiat_eff}.

\subsection{Origin of the second bright episode}
\label{sec:second_episode_origin}

The second bright episode exhibits a pronounced
GBM and BAT pulse accompanied by an XRT flare
(Figure~\ref{fig:second_pulse}). Their temporal
association suggests an additional emission
component superposed on the underlying afterglow,
although it does not establish a common emission
mechanism across these bands.
Internal dissipation through internal shocks or
magnetic reconnection provides one possible
explanation \citep{1994ApJ...430L..93R,2011ApJ...726...90Z}.

The energy-dependent pulse profiles and hardness
evolution provide complementary constraints.
GBM and XRT show opposite average hardness trends
over the covered episode, whereas the XRT hardness
is consistent with remaining constant during
the observed decline.
These differences do not by themselves distinguish
internal dissipation from an external contribution.

LAT emission is detected during 332.8--390~s,
but the five-bin light curve does not resolve
a distinct pulse comparable to those observed
at lower energies.
Within our afterglow model, the LAT emission
can remain dominated by the external-shock
component. The interval-averaged measurement
nevertheless allows a shorter or weaker
contribution associated with the second episode.

The GBM spectrum is described by an empirical
Comptonized function
(Table~\ref{tab:gbm_spectral_parameters}),
which does not identify a specific radiation
mechanism.
Likewise, the XRT decay analysis in Section~\ref{sec:second_pulse}
does not uniquely determine the emission origin:
the fitted temporal index depends strongly on
the assumed time origin, and coverage ends at
359.37~s, before the full flare decay is observed.
These limitations prevent a decisive
curvature-effect test
\citep{2006ApJ...646..351L}.

\subsection{Multiwavelength comparison with the GRB
population}
\label{sec:population_comparison}

Figures~\ref{fig:x-ray_luminosity}--\ref{fig:LAT_comp} place GRB~260924A in the context
of other GRBs across the X-ray, optical, and GeV
bands. These comparisons use different observables:
isotropic-equivalent rest-frame X-ray luminosities,
optical magnitudes transformed to a common $z=1$
system, and observed LAT photon fluxes.

In X-rays, GRB~260924A lies near the high-luminosity
end of the \textit{Swift}/XRT comparison sample
(Figure~\ref{fig:x-ray_luminosity}). Table~\ref{tab:xray_luminosity_ranking} quantifies its position
at selected rest-frame epochs. At 3500, 7200,
and 20400~s, its central luminosity estimate is
the second highest among the 75, 62, and 35
events, respectively, satisfying the common-epoch
coverage criteria. These nominal ranks do not
establish statistically significant differences
between neighbouring events, and the quoted
luminosity uncertainties include statistical
errors only.  At a rest-frame time of 3500~s, its inferred
luminosity is comparable to that of GRB~221009A:
the statistical uncertainties overlap, while
systematic differences between the spectral
and extraction procedures have not been quantified. The nominal
first-place position at 75~s occurs during
the flare associated with the second bright
episode and depends on the adopted mode-averaged
spectral conversion; it does not characterize
the underlying afterglow.

The optical afterglow is similarly remarkable.
Figure~\ref{fig:kann} compares its evolution with the
long-GRB sample of \citet{Kann2006,Kann2010,Kann2011,Kann2024} and with GRB~221009A using the
observations of \citet{OConnor2023}.
After transforming the light curves to
$R$-band AB magnitudes in a common $z=1$
system, GRB~260924A lies near the upper
envelope of the sample out to approximately
one day.
In this reference frame, it is approximately
2~mag brighter than GRB~221009A at 0.3~d
and 1~mag brighter near 1~d, corresponding
to flux ratios of approximately 6.3 and 2.5.
These differences depend on the extinction
and spectral corrections applied to both events.

In the GeV band, GRB~260924A lies towards the
bright end of the selected LAT comparison
sample over approximately 70--1100~s,
particularly during its first observed interval
(Figure~\ref{fig:LAT_comp}). The comparison uses observer-frame times and
photon fluxes integrated above 100~MeV, with
dataset-dependent upper energy limits of
100, 300, or 1000~GeV. For GRB~260924A,
we use the 0.1--1000~GeV band.
It therefore characterizes apparent brightness
and temporal evolution without establishing
a ranking in intrinsic GeV luminosity.

Together with the prompt energetics discussed
in Section~\ref{radiat_eff}, these comparisons show that
GRB~260924A combines an extreme
isotropic-equivalent gamma-ray energy with
luminous X-ray and optical afterglows and
prominent observed GeV emission.
The broadband prominence need not arise from
a single radiation mechanism: the afterglow
luminosities also depend on the circumburst
environment and shock microphysics.

\section{Conclusions}
\label{sec:conclusions}

Our multiwavelength study places GRB~260924A
among the most energetic long GRBs observed.
At $z=3.51$, its isotropic-equivalent prompt
gamma-ray energy is $E_{\gamma,\mathrm{iso}}\simeq9.79\times10^{54}$~erg.
The prompt emission comprises two bright episodes,
the second beginning near 333~s, separated by
weaker, energy-dependent activity.
BAT and LAT detect emission between the episodes,
including a 14.2~GeV photon at 304~s.
The highest-energy photon associated with the
burst, at 43.3~GeV, arrived at 82~s during
the first episode.

LAT emission extends to approximately 1100~s.
The four intervals after 164.9~s follow a decay
with $\alpha_{\rm LAT}=1.30^{+0.20}_{-0.19}$,
whose backward extrapolation underpredicts
the first-interval average by a factor of
$2.25\pm0.59$.
The five interval-resolved spectra are compatible
with a shared photon index,
$\Gamma_{\rm LAT}=2.24\pm0.06$.
The second bright GBM/BAT episode is accompanied
by an X-ray flare, whereas the LAT light curve
does not resolve a corresponding pulse.
The incomplete XRT coverage of the flare decay
limits constraints on its physical origin.

The X-ray and optical afterglows lie near
the high-luminosity end of their respective
comparison samples.
In the common $z=1$ system, the optical afterglow
is approximately 2~mag brighter than GRB~221009A
at 0.3~d and 1~mag brighter near 1~d under
the adopted corrections.
Its rest-frame 0.3--10~keV X-ray luminosity
is among the highest at the selected post-flare
epochs and is comparable to that of
GRB~221009A at a rest-frame time of 3500~s.
The LAT comparison likewise shows prominent
observed GeV emission, without establishing
an intrinsic luminosity ranking.

Under the single-synchrotron-segment working
hypothesis, the late afterglow observations favour
a stratified circumburst medium, which we model
as $n(R)\propto R^{-1.2}$.  We describe the broadband light curves through
forward-shock synchrotron and synchrotron
self-Compton emission, with reverse-shock SSC
accounting for the early LAT excess.
The highest-energy LAT photons exceed the
predicted synchrotron energies under the adopted
acceleration prescriptions, further supporting
an SSC contribution.
The later X-ray rebrightening is described
through energy injection from trailing ejecta
and evolution of the electron spectral index
and microphysical energy fractions during
4700--6200~s.
The kinetic energy and these parameters remain
constant afterwards.
This light-curve description does not resolve
the early X-ray spectral discrepancy.

The inferred pre-injection isotropic-equivalent
kinetic energy,
$E_{\mathrm{K,iso}}\simeq7.0\times10^{55}$~erg,
corresponds to a prompt radiative efficiency
of $\eta_\gamma\simeq0.12$.
Within the adopted model, the extreme prompt
energy is accompanied by a large kinetic-energy
reservoir without requiring an unusually high
radiative efficiency.
The magnetic contrast $R_B\simeq2.6$ inferred
from the combined shock model is compatible
with a magnetic field retained by the ejecta,
but does not directly measure their upstream
magnetization.
The available optical and X-ray observations
do not establish a robust jet-break epoch,
leaving the jet opening angle and
beaming-corrected energy undetermined.

GRB~260924A thus provides a valuable case
for examining how a large relativistic
energy reservoir, the circumburst environment,
and shock microphysics jointly shape
the broadband afterglow.

%%%%%%%%%%%%%%%%%%%%%%%%%%%%%%%%%%%%%%%%%%%%%%%%%%

\section*{Acknowledgements}

NF is grateful to UNAM-DGAPA-PAPIIT for the funding provided by grant IN112525. 
MGD acknowledges the support of the DoS and by JSPS Grant-in-Aid for Scientific Research (KAKENHI) (A), Grant Number JP25H00675. 

%%%%%%%%%%%%%%%%%%%%%%%%%%%%%%%%%%%%%%%%%%%%%%%%%%
\section*{DATA AVAILABILITY}

The \textit{Fermi}/GBM and LAT and \textit{Swift}/BAT, XRT, and UVOT observations
analysed in this work are available from the respective public
mission archives. The derived light curves, interval definitions,
fit summaries, and scripts used to produce the figures are available
from the corresponding author upon reasonable request. 
%The analysis products will be deposited in a public repository before publication; the repository identifier will be added to the accepted manuscript.


\bibliographystyle{mnras}
\bibliography{references_coc} % if your bibtex file is called example.bib

\clearpage
\newpage

\begin{table*}
\centering
\caption{GBM spectral parameters for the selected intervals
of GRB~260924A, using linear temporal backgrounds.
Alternative continuum and candidate Band+BB fits
are listed for comparison.}
\label{tab:gbm_spectral_parameters}
\begingroup
\small
\setlength{\tabcolsep}{3.5pt}
\renewcommand{\arraystretch}{1.3}
\begin{tabular}{@{}llccccc rr@{}}
\toprule
Interval & Model & $\alpha$ & $\beta$ &
$E_{\rm peak}$ & $k_{\rm B}T$ & PG &
$N_{\rm ch}$ & $N_{\rm src}$\\
(s) & & & & (keV) & (keV) & & & \\
\midrule
$-6.1$--$164.9$ & Band
& $-0.85^{+0.01}_{-0.01}$
& $-3.52^{+0.35}_{-0.64}$
& $596^{+12}_{-12}$ & --
& $687.25$ & 485 & 4\\
& Band+BB
& $-0.91^{+0.02}_{-0.02}$
& $-3.79^{+0.46}_{-1.59}$
& $681^{+27}_{-26}$
& $42.7^{+3.5}_{-3.4}$
& $666.31$ & 485 & 6\\
\addlinespace
$-6.1$--$50^{a}$ & Band
& $-0.65$ & $-12^{b}$ & $350.7$ & --
& $524.22$ & 485 & 4\\
& Band+BB
& $-0.77$ & $-12^{b}$ & $453.9$ & $33.10$
& $507.30$ & 485 & 6\\
\addlinespace
$50$--$115$ & Band
& $-0.87^{+0.01}_{-0.01}$
& $-3.16^{+0.19}_{-0.27}$
& $764^{+18}_{-17}$ & --
& $715.46$ & 485 & 4\\
& Band+BB
& $-0.91^{+0.02}_{-0.02}$
& $-3.27^{+0.23}_{-0.35}$
& $836^{+37}_{-34}$
& $60.2^{+9.9}_{-8.7}$
& $708.59$ & 485 & 6\\
\addlinespace
$115$--$164.9$ & Band
& $-0.89^{+0.02}_{-0.01}$
& $-4.15^{c}$
& $504^{+16}_{-17}$ & --
& $501.52$ & 485 & 4\\
\addlinespace
$332.8$--$390$ & Comptonized
& $-1.05^{+0.03}_{-0.03}$ & --
& $714^{+64}_{-55}$ & --
& $355.92$ & 391 & 3\\
& Band
& $-1.03^{+0.03}_{-0.03}$
& $-2.15^{d}$
& $617^{+73}_{-72}$ & --
& $349.15$ & 391 & 4\\
\bottomrule
\end{tabular}
\endgroup

\vspace{2mm}
\begin{minipage}{\textwidth}
\footnotesize
Times are relative to the GBM trigger.
The integrated first-episode spectrum and its
subintervals use overlapping data and are not
independent measurements.
Reported uncertainties correspond to nominal
68.27 per cent one-parameter profile intervals
at $\Delta{\rm PG}=1$, conditional on the adopted
model, background, and channel selection.
${\rm PG}=2S$, where $S=-\ln\mathcal{L}$ up to
data-dependent constants.
$N_{\rm ch}$ and $N_{\rm src}$ give the numbers of
fitted channels and free source parameters,
respectively; background nuisance parameters are
excluded from $N_{\rm src}$.
These counts do not define calibrated degrees of freedom.\\
$^{a}$ Profile uncertainties are unavailable in the
reviewed archive; point estimates are reported.\\
$^{b}$ The optimum reaches the imposed lower bound
$\beta=-12$.\\
$^{c}$ The $\Delta{\rm PG}=1$ profile support is
$[-12,-3.12]$, reaching the lower search boundary.\\
$^{d}$ The corresponding support is
$[-2.43,-2.01]$, reaching the upper search boundary.
Search-boundary endpoints are not statistical
confidence limits.
\end{minipage}
\end{table*}

\begin{table*}
\centering
\caption{Continuum and BB normalizations and
within-interval model comparisons for the linear-background
GBM fits in Table~\ref{tab:gbm_spectral_parameters}.}
\label{tab:gbm_spectral_normalizations}
\begingroup
\small
\setlength{\tabcolsep}{7pt}
\renewcommand{\arraystretch}{1.3}
\begin{tabular}{@{}llccc@{}}
\toprule
Interval (s) & Model
& $A/(10^{-2}\,\mathrm{ph\,cm^{-2}\,s^{-1}\,keV^{-1}})$
& $A_6$ & $\Delta\mathrm{PG}$\\
\midrule
$-6.1$--$164.9$ & Band
& $1.810^{+0.017}_{-0.016}$ & -- & --\\
 & Band+BB
& $1.624^{+0.043}_{-0.042}$
& $1.33^{+0.39}_{-0.34}$ & $20.94$\\
\addlinespace
$-6.1$--$50^{a}$ & Band & $1.19$ & -- & --\\
 & Band+BB & $0.85$ & $3.8$ & $16.93$\\
\addlinespace
$50$--$115$ & Band
& $2.42^{+0.02}_{-0.02}$ & -- & --\\
 & Band+BB
& $2.26^{+0.06}_{-0.06}$
& $0.47^{+0.26}_{-0.21}$ & $6.88$\\
\addlinespace
$115$--$164.9$ & Band
& $1.88^{+0.03}_{-0.03}$ & -- & --\\
\addlinespace
$332.8$--$390$ & Comptonized
& $0.82^{+0.02}_{-0.02}$ & -- & --\\
 & Band
& $0.85^{+0.03}_{-0.03}$ & -- & $6.77$\\
\bottomrule
\end{tabular}
\endgroup

\vspace{2mm}
\begin{minipage}{\textwidth}
\footnotesize
$A$ is the continuum prefactor at the fixed
100~keV pivot.  For the BB prescription defined in Appendix~\ref{app:gbm_blackbody}, $A_{\rm BB}=10^{-6}A_6\,
\mathrm{ph\,cm^{-2}\,s^{-1}\,keV^{-3}}$. Normalizations refer to the indicated joint fit.
Time and uncertainty conventions, including
superscript $a$, follow
Table~\ref{tab:gbm_spectral_parameters}.
For Band+BB,
$\Delta{\rm PG}={\rm PG}_{\rm Band}
-{\rm PG}_{\rm Band+BB}$;
for the second-episode Band alternative,
$\Delta{\rm PG}={\rm PG}_{\rm Comp}
-{\rm PG}_{\rm Band}$.
These differences measure fit improvements,
not detection significances.
\end{minipage}
\end{table*}

\begin{table*}
\centering
\caption{Time-resolved LAT spectral and flux measurements
of GRB~260924A in 0.1--1000~GeV.}
\label{tab:lat_integrated_spectra}
\begingroup
\small
\setlength{\tabcolsep}{8pt}
\renewcommand{\arraystretch}{1.2}
\begin{tabular}{@{}ccccc@{}}
\toprule
Interval & $\Gamma_{\rm LAT}$ & $F_{\rm ph}$ & $F_E$ & TS \\
(s) & &
($10^{-5}$ ph cm$^{-2}$ s$^{-1}$) &
($10^{-8}$ erg cm$^{-2}$ s$^{-1}$) & \\
\midrule
70.04--164.9  & $2.21\pm0.08$ & $60.25\pm4.38$ & $47.43\pm8.95$ & 1058.36 \\
164.9--332.8  & $2.24\pm0.14$ & $10.01\pm1.23$ & $7.40\pm2.35$  & 282.41 \\
332.8--390    & $2.23\pm0.34$ & $5.06\pm1.48$  & $3.78\pm2.96$  & 63.55 \\
390--800.53   & $2.28\pm0.15$ & $3.20\pm0.43$  & $2.17\pm0.67$  & 309.28 \\
800.53--1100  & $2.52\pm0.45$ & $2.00\pm0.72$  & $0.93\pm0.52$  & 26.95 \\
\bottomrule
\end{tabular}
\endgroup

\vspace{2mm}
\begin{minipage}{\textwidth}
\footnotesize
Times are relative to the GBM trigger.
$F_{\rm ph}$ and $F_E$ are interval-averaged photon
and energy fluxes, respectively.
Errors are nominal $1\sigma$ uncertainties from the
local covariance matrix; energy-flux errors include
normalization--index covariance.
Both diffuse normalizations are free within $[0,30]$
under the source and background-only hypotheses.
Sensitivity tests are described in Appendix~\ref{app:lat_systematics}.
\end{minipage}
\end{table*}

\begin{table*}
\centering
\renewcommand{\arraystretch}{1.3}\addtolength{\tabcolsep}{1.8pt}
\caption{The ten highest-energy LAT events with
$P_{\rm GRB}\geq0.9$ in the full-window fit. Arrival times
are relative to the GBM trigger, and separations are measured
from the adopted XRT position. Association probabilities
depend on the fitted LAT model.}
\label{tab:lat_high_energy_events}
\begin{tabular}{rrrr}
\hline
$E_{\rm obs}$ & $t-t_{\rm GBM,0}$ & Separation & $P_{\rm GRB}$ \\
(GeV) & (s) & (deg) & \\
\hline
43.31 &  82.24 & 0.070 & 0.999992 \\
24.95 & 122.92 & 0.41 & 0.998767 \\
14.24 & 304.15 & 0.08 & 0.999993 \\
11.59 & 141.64 & 0.18 & 0.999957 \\
 5.30 & 665.55 & 0.05 & 0.999996 \\
 3.95 &  98.99 & 0.28 & 0.999857 \\
 3.52 &  88.12 & 0.61 & 0.999316 \\
 3.11 & 587.17 & 0.11 & 0.999985 \\
 2.63 & 136.79 & 0.24 & 0.999920 \\
 2.36 & 147.86 & 0.23 & 0.999937 \\
\hline
\end{tabular}
\end{table*}

\begin{table*}
\centering
\renewcommand{\arraystretch}{1.5}\addtolength{\tabcolsep}{2pt}
\caption{BAT power-law fits in 15-150\,keV. Times are relative
to the GBM trigger. Fluxes are energy fluxes; uncertainties
are 90 per cent confidence intervals for one parameter.}
\label{tab:bat_pl}
\begin{tabular}{lccc}
\hline
Interval (s) &
$\Gamma_{\rm BAT}$ &
$F_{15-150}$ &
$\chi^2/{\rm dof}$ \\
&
&
$(10^{-7}\,\mathrm{erg\,cm^{-2}\,s^{-1}})$
& \\
\hline
70-165  &
$1.08^{+0.02}_{-0.02}$ &
$3.14\pm0.04$ &
$36.30/56$ \\
250-320 &
$1.55^{+0.28}_{-0.28}$ &
$0.08^{+0.01}_{-0.01}$ &
$46.40/56$ \\
334-390 &
$1.16^{+0.03}_{-0.04}$ &
$1.48\pm0.03$ &
$53.49/56$ \\
\hline
\end{tabular}
\end{table*}

\begin{table*}
\renewcommand{\arraystretch}{1.5}\addtolength{\tabcolsep}{2pt}
\centering
\caption{Automated \textit{Swift}/XRT temporal fit from the UKSSDC
catalogue \citep{Evans2009}.
The five-segment model follows $F_X\propto t^{-\alpha_i}$
and gives $\chi^2/\mathrm{dof}=293/336$.
Times are relative to the BAT trigger;
uncertainties are quoted at 90 per cent confidence.}
\label{tab:xrt_catalogue_temporal}
\begin{tabular}{ccc}
\hline
Segment & $\alpha_i$ & Break after segment, $t_{b,i}$ (s; BAT $T_0$) \\
\hline
1 & $0.3^{+0.5}_{-0.7}$       & $130^{+8}_{-6}$ \\
2 & $2.85^{+0.12}_{-0.12}$    & $256.4^{+15.7}_{-11.1}$ \\
3 & $1.316^{+0.025}_{-0.029}$ & $4.67^{+0.22}_{-0.44}\times10^3$ \\
4 & $-1.50^{a}$        & $6.30^{+4.87}_{-0.26}\times10^3$ \\
5 & $1.61^{+0.05}_{-0.05}$   & --- \\\hline
\end{tabular}
\par\vspace{3pt}
\begin{minipage}{\textwidth}
\footnotesize
\textit{Notes.} The catalogue excludes 115--127 and 311--3647~s
from the temporal fit; these windows are not
measurements of flare duration.
The fourth break lies within a coverage gap
and is inferred from the fitted segments.
The GBM reference is obtained using
$t_{\rm GBM}=t_{\rm BAT}+4.71$~s.\\
$^{a}$ The catalogue displays the uncertainty on
$\alpha_4$ as $(+0.00,+2.15)$.
We retain the best-fitting value without
interpreting these entries as a conventional
two-sided confidence interval.
\end{minipage}
\end{table*}

\begin{table*}
\renewcommand{\arraystretch}{1.5}\addtolength{\tabcolsep}{2pt}
\centering
\caption{Joint spectral analysis of the early \textit{Swift}/XRT emission of GRB 260924A.}
\label{tab:xrt_common_absorption}
\begin{tabular}{ccccc}
\hline\hline
Mode &
Time since GBM trigger & Exposure & $\Gamma_{\rm XRT}$ & $F_{\nu,\rm unabs}(1\,{\rm keV})$ \\ & (s) & (s) & & ($\mu$Jy) \\
\hline
WT & 263.5--314.0   & 50.45  & $1.78^{+0.07}_{-0.07}$ & $631^{+38}_{-36}$ \\
WT & 3652.5--3757.0 & 104.36 & $2.25^{+0.15}_{-0.14}$ & $26.0^{+2.4}_{-2.2}$ \\
PC & 3763.4--4656.0 & 889.03 & $2.01^{+0.10}_{-0.10}$ & $18.4^{+1.4}_{-1.3}$ \\
\hline
\end{tabular}

\medskip
\begin{minipage}{0.96\textwidth}
\footnotesize
\textit{Notes.}
The spectra were fitted jointly over 0.3--10~keV
with \texttt{TBabs*zTBabs*powerlaw}.
The Galactic column was fixed at
$N_{\rm H,Gal}=4.09\times10^{20}$~cm$^{-2}$.
The shared intrinsic column at $z=3.51$ is
$N_{\rm H,int}=(7.46^{+5.17}_{-4.88})
\times10^{21}$~cm$^{-2}$;
photon indices and normalizations vary independently.
The joint fit gives $W=896.40$ for 1009 degrees
of freedom.
Errors correspond to one-parameter 90 per cent
profile intervals, $\Delta W=2.71$.
Flux-density uncertainties follow from the profiled
power-law normalizations.
Time intervals delimit the selected GTIs;
exposures give the live integration times.
\end{minipage}
\end{table*}

\begin{table*}
\centering
\caption{Count-rate hardness ratios of the second bright episode shown
in Fig.~\ref{fig:second_pulse}. Each instrument retains its own time bins; entries sharing a
row are not necessarily simultaneous measurements.}
\label{tab:second_pulse_hardness}
\small
\setlength{\tabcolsep}{4pt}
\renewcommand{\arraystretch}{1.2}
\begin{tabular}{cccccc}
\hline
\multicolumn{2}{c}{\textit{Fermi}/GBM} & \multicolumn{2}{c}{\textit{Swift}/BAT} & \multicolumn{2}{c}{\textit{Swift}/XRT} \\
Interval (s) & $H_{\rm GBM}$ & Interval (s) & $H_{\rm BAT}$ & Interval (s) & $H_{\rm XRT}$ \\
\hline
334-338 & $1.32^{+0.30}_{-0.22}$ & 333.93-337.93 & $0.96^{+0.16}_{-0.13}$ & 321.11-324.52 & $0.63\pm0.12$ \\
338-342 & $1.85^{+0.12}_{-0.11}$ & 337.93-341.93 & $1.06^{+0.07}_{-0.06}$ & 324.52-328.53 & $0.96\pm0.18$ \\
342-346 & $1.75^{+0.04}_{-0.04}$ & 341.93-345.93 & $1.05^{+0.03}_{-0.03}$ & 328.53-331.67 & $0.99\pm0.18$ \\
346-350 & $1.35^{+0.05}_{-0.05}$ & 345.93-349.93 & $0.95^{+0.03}_{-0.03}$ & 331.67-334.39 & $1.03\pm0.18$ \\
350-354 & $1.28^{+0.07}_{-0.07}$ & 349.93-353.93 & $0.93^{+0.04}_{-0.04}$ & 334.38-337.04 & $0.94\pm0.17$ \\
354-358 & $1.06^{+0.09}_{-0.08}$ & 353.93-357.93 & $0.86^{+0.05}_{-0.05}$ & 337.04-338.76 & $1.37\pm0.26$ \\
358-362 & $0.77^{+0.09}_{-0.08}$ & 357.93-361.93 & $0.81^{+0.08}_{-0.07}$ & 338.76-340.46 & $1.69\pm0.26$ \\
362-366 & $0.75^{+0.10}_{-0.09}$ & 361.93-365.93 & $0.82^{+0.10}_{-0.09}$ & 340.46-342.54 & $1.63\pm0.27$ \\
366-370 & $0.93^{+0.22}_{-0.17}$ & 365.93-369.93 & $0.96^{+0.16}_{-0.13}$ & 342.54-344.35 & $1.90\pm0.29$ \\
370-374 & $1.14^{+0.43}_{-0.28}$ & 369.93-373.93 & $0.57^{+0.12}_{-0.11}$ & 344.35-345.57 & $1.77\pm0.29$ \\
374-378 & $0.65^{+0.43}_{-0.27}$ & 373.93-377.93 & $1.22^{+0.46}_{-0.31}$ & 345.57-347.13 & $1.77\pm0.28$ \\
378-382 & $0.38^{+0.35}_{-0.25}$ & 377.93-381.93 & $0.54^{+0.16}_{-0.14}$ & 347.13-349.11 & $1.98\pm0.30$ \\
382-386 & $0.84^{+0.69}_{-0.35}$ & 381.93-385.93 & $1.17^{+0.41}_{-0.28}$ & 349.11-351.22 & $1.66\pm0.27$ \\
386-390 & $0.62^{+1.23}_{-0.45}$ & 385.93-389.93 & $1.20^{+1.11}_{-0.50}$ & 351.22-354.01 & $2.28\pm0.35$ \\
-- & -- & -- & -- & 354.01-356.61 & $1.78\pm0.29$ \\
-- & -- & -- & -- & 356.61-360.10 & $1.62\pm0.22$ \\
\hline
\end{tabular}
\par\smallskip
\begin{minipage}{\textwidth}
\footnotesize
\textit{Notes.} Hardness ratios are defined as
$H_{\rm GBM}=C_{50-300}/C_{8-50}$,
$H_{\rm BAT}=C_{50-100}/C_{25-50}$, and
$H_{\rm XRT}=C_{1.5-10}/C_{0.3-1.5}$,
where $C$ denotes count rate and energy bands
are in keV.
GBM combines Na\,I detectors n3, n4, and n5.
GBM and BAT use 4-s groups of native bins,
retaining their background treatment without
additional local-level subtraction.
Their errors are nominal $\Delta\chi^2=1$
Gaussian profile intervals.
XRT values, bins, and errors follow the native
Swift Burst Analyser product.
Times are relative to the GBM trigger.
Different energy bands and instrumental responses
prevent direct comparison of hardness amplitudes
between instruments.
\end{minipage}
\end{table*}

\begin{table*}
\centering
\caption{Optical/NIR and optical-to-X-ray spectral
indices for the three SED epochs.}
\label{tab:optical_xrt_sed}
\begin{tabular}{cccc}
\hline
Interval & $\beta_{\rm opt/NIR}$ &
$\beta_{\rm opt/NIR/X}$ & $\chi^2/\mathrm{dof}$ \\
(d) & & & \\
\hline
0.18-0.20 & $0.57\pm0.07$ & $0.85\pm0.01$ & 17.41/1 \\
0.85-1.00 & $0.71\pm0.06$ & $0.89\pm0.01$ & 8.80/1 \\
2.30-2.70 & $0.68\pm0.10$ & $0.90\pm0.01$ & 4.70/2 \\
\hline
\end{tabular}
\par\smallskip
\begin{minipage}{\textwidth}
\footnotesize
\textit{Notes.} Times are relative to the GBM trigger.
Spectral indices follow $F_\nu\propto\nu^{-\beta}$.
The optical/NIR-only indices are listed separately
from the single-power-law diagnostics including XRT;
the last column refers to the latter.
Errors are formal $1\sigma$ uncertainties, conditional
on the adopted extinction corrections and fixed XRT
spectral conversion.
All three XRT flux densities are predictions from
the same temporal fit and are correlated; the latter
two epochs lie in XRT coverage gaps.
The first two optical/NIR-only slopes are two-point
colours and do not independently test a power-law
spectral shape.
\end{minipage}
\end{table*}

\begin{table*}
\centering
\caption{X-ray luminosity of GRB~260924A and its nominal
position within the comparison sample at selected
rest-frame epochs. Luminosities refer to the
absorption-corrected rest-frame 0.3--10~keV band.}
\label{tab:xray_luminosity_ranking}
\small
\setlength{\tabcolsep}{4pt}
\renewcommand{\arraystretch}{1.2}
\begin{tabular}{@{}r l c c p{4.4cm}@{}}
\hline
$t_{\rm rest}$ & Phase of GRB~260924A
& $L_X$ & Rank/$N$\textsuperscript{a}
& Brighter GRBs\textsuperscript{b} \\
(s) & & (erg\,s$^{-1}$) & & \\
\hline
60
& Early clean interval
& $5.26^{+0.60}_{-0.50}\times10^{50}$
& $8/191$
& 060526, 181110A, 100302A, 140114A,
  130514A, 090516A, 241030A \\

75
& Flare\textsuperscript{c}
& $6.71^{+1.45}_{-0.86}\times10^{50}$
& $1/206$
& None \\

820
& Resumed WT coverage
& $(3.52\pm0.58)\times10^{49}$
& $3/70$
& 150206A, 220101A \\

1000
& Resumed PC coverage
& $(2.27\pm0.42)\times10^{49}$
& $2/69$
& 050730 \\

3500
& Later decay
& $(4.81\pm0.89)\times10^{48}$
& $2/75$
& 240529A \\

7200
& Later decay
& $(2.37\pm0.38)\times10^{48}$
& $2/62$
& 111209A \\

20400
& Later decay
& $(3.75\pm0.89)\times10^{47}$
& $2/35$
& 220101A \\
\hline
\end{tabular}

\vspace{2mm}
\begin{minipage}{\textwidth}
\footnotesize
\textsuperscript{a}
$N$ includes GRB~260924A and varies with the available
coverage at each epoch. One luminosity estimate per
event is obtained by local interpolation in
$\log t$--$\log L_X$, requiring a separation between
bracketing bin centres of at most 0.10~dex, an
inter-bin gap no larger than 5\% of the comparison
epoch, and individual bin widths no larger than 20\%
of that epoch. No extrapolation is performed.
Stricter coverage criteria preserve the nominal ranks.

\textsuperscript{b}
Ranks compare central luminosity estimates and do not
establish statistically significant differences
between neighbouring events. Quoted uncertainties
include statistical errors only. Flares are retained
in the comparison curves. Excluding GRB~140114A
(approximate redshift) and GRB~130514A
(photometric redshift) changes the 60~s rank from
eighth to sixth. Also excluding the less secure
host associations of GRBs~181110A and 241030A changes
it to fourth. The other listed ranks are unchanged.

\textsuperscript{c}
The flare luminosity uses the mode-averaged spectral
conversion. Its nominal first-place position is
conditional on this prescription and does not
characterize the underlying afterglow.
\end{minipage}
\end{table*}

\begin{table*}
\centering
\renewcommand{\arraystretch}{1.2}
\setlength{\tabcolsep}{15pt} %
\caption{Posterior medians and 16th--84th percentile intervals
for the forward-shock and combined forward- and
reverse-shock models of GRB~260924A.  Prior bounds and posterior values use the scale factors indicated in the first column.
$U(a,b)$ denotes a uniform distribution between $a$ and $b$. }
\label{tab5:fit}
%\resizebox{0.3\columnwidth}{!}{%
\begin{tabular}{l c c c }
\hline
\hline
Parameter & Priors & Values  & Values   \\ \hline \hline

Shock: &  & \bf{Forward}          & \bf{Forward + Reverse }\\
& & --------------------------    & -------------------------- \\
${\rm E_{\rm K,iso}\,(10^{55}\,{\rm erg})}$ & $U(3.0, 9.16)$  & $7.02^{+0.41}_{-0.40}$   & $7.01^{+0.39}_{-0.37}$  \\
${\rm n_{\rm dec}\,(cm^{-3})}$  & $U(0.8, 7.2)$ & $2.46_{-0.57}^{+0.58}$ & $2.51_{-0.46}^{+0.51}$ \\
${\rm p}$               &  $U(2.2, 2.5)$ & $2.37_{-0.06}^{+0.06}$ & $2.35_{-0.07}^{+0.07}$ \\
${ \rm log_{10} [\Gamma_0}]$  & $U(1, 3.1)$ & $2.57_{-0.28}^{+0.27}$ & $2.61_{-0.29}^{+0.28}$  \\
${\varepsilon_{\rm B_f}(10^{-5})}$  & $U(1.0, 6.1)$ & $1.50_{-0.34}^{+0.34}$  & $1.42_{-0.37}^{+0.39}$  \\
${\varepsilon_{\rm e_f}}(10^{-2})$  &  $U(0.5, 3.0)$  & $2.06_{-0.20}^{+0.20}$ & $2.11_{-0.18}^{+0.19}$  \\
${\rm \alpha_B}$              & $U(0.92, 3.5)$  & $3.00_{-0.36}^{+0.34}$   & $3.00_{-0.36}^{+0.34}$ \\
${\rm \alpha_e}$              & $U(1.14, 10.5)$   & $3.89_{-0.54}^{+0.54}$  & $3.89_{-0.54}^{+0.54}$ \\
${\zeta_{\rm e_f}}$  & $U(0.1, 1)$  & $0.88_{-0.10}^{+0.10}$  & $0.89_{-0.11}^{+0.11}$\\

${\varepsilon_{\rm B_r}(10^{-5})}$ & $U(4.85, 12.1)$ &  $-$  & $9.78_{-0.34}^{+0.34}$   \\
${\varepsilon_{\rm e_r}}(10^{-1})$ & $U(1.5, 4.19)$ &  $-$  & $2.04_{-0.20}^{+0.21}$  \\
${\zeta_{\rm e_r}}\,(10^{-2})$    & $U(2.8, 4)$  &  $-$  & $3.42_{-0.11}^{+0.11}$  \\
\hline

\end{tabular}%
%}
\end{table*}

\newpage
\clearpage

\begin{figure*}
\centering
\includegraphics[width=0.8\textwidth]{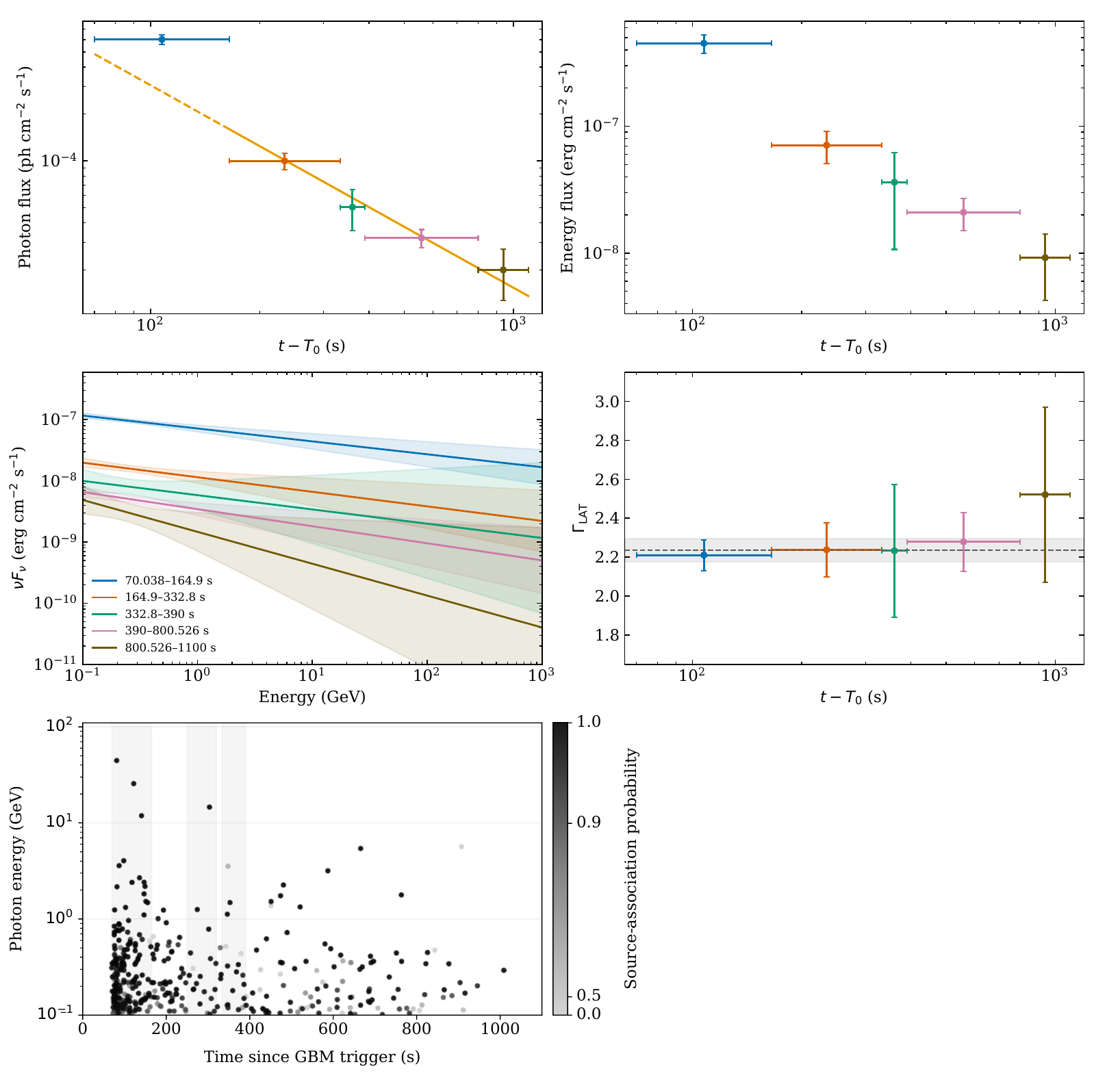}
\caption{\textit{Fermi}/LAT temporal and spectral evolution of
GRB~260924A.
Top left: photon flux in 0.1--300~GeV;
the orange curve shows the power-law fit to the
four later intervals and its backward extrapolation,
converted from the fitted 0.1--1000~GeV band.
Top right: energy flux in 0.1--300~GeV.
Middle left: fitted power-law spectra with
uncertainty envelopes.
Middle right: photon indices; the horizontal
band shows the shared-index fit and its uncertainty.
Bottom: energies and arrival times of all
404 selected events, with greyscale indicating
their model-dependent source-association probabilities.
Shading marks the bright BAT episodes.
Horizontal bars span the spectral intervals;
vertical bars indicate nominal $1\sigma$
uncertainties.
Times are relative to the GBM trigger.}
\label{fig:LAT_lightcurve}
\end{figure*}

\begin{figure*}
 \centering
\includegraphics[width=0.7\linewidth]{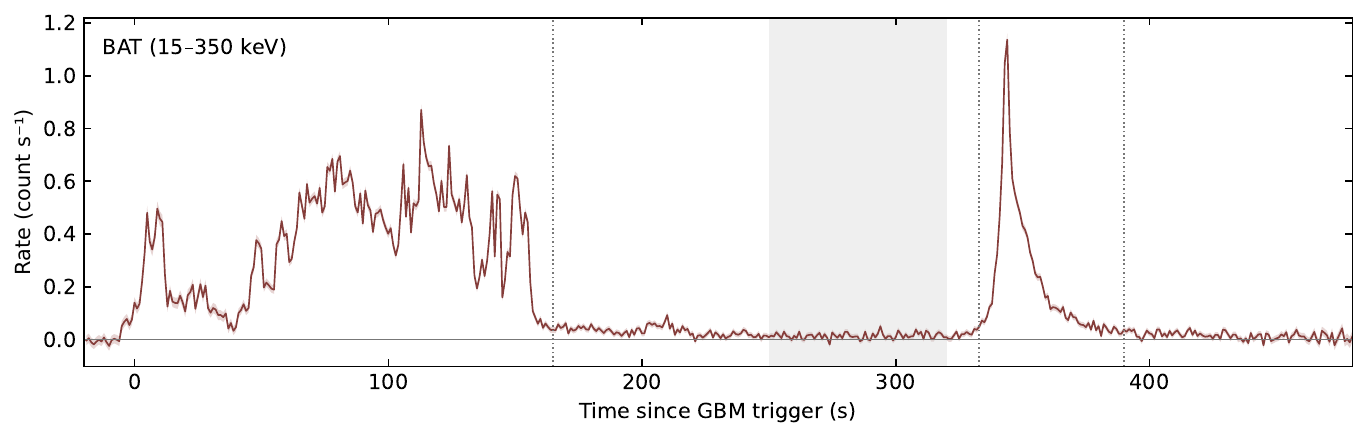} 
 \includegraphics[width=0.7\linewidth]{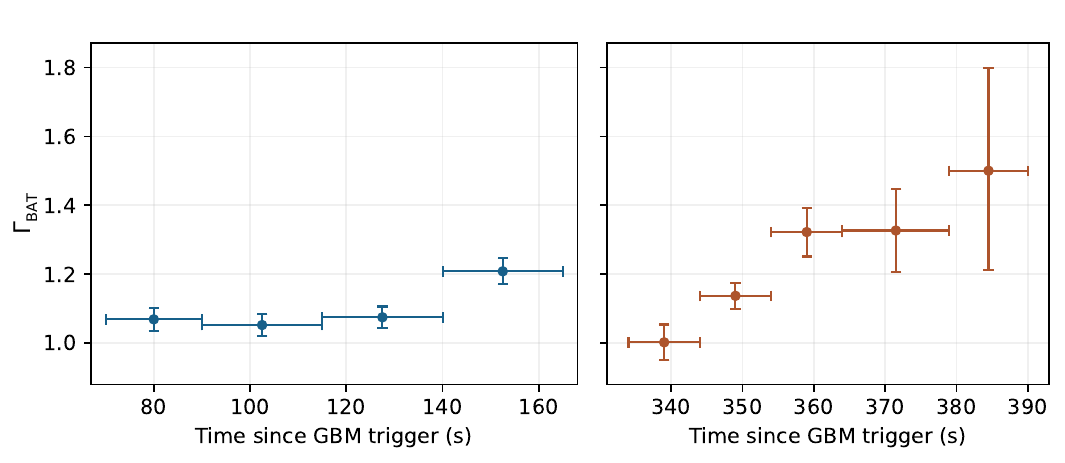} 
\caption{\textit{Swift}/BAT temporal and spectral evolution of
GRB~260924A. The upper panel shows the mask-weighted count-rate
light curve in 15-350~keV. The shaded region marks the
250-320~s inter-episode spectral interval. The lower panels show the
photon index obtained from power-law fits in 15-150~keV
during the initial bright episode and the second pulse.
Horizontal bars indicate the spectral extraction intervals;
vertical bars show the reported 90 per cent uncertainties
for one parameter. All times are relative to the GBM trigger.}
\label{fig:BAT_photon}
\end{figure*}

\begin{figure*}
\centering
\includegraphics[width=0.6\linewidth]{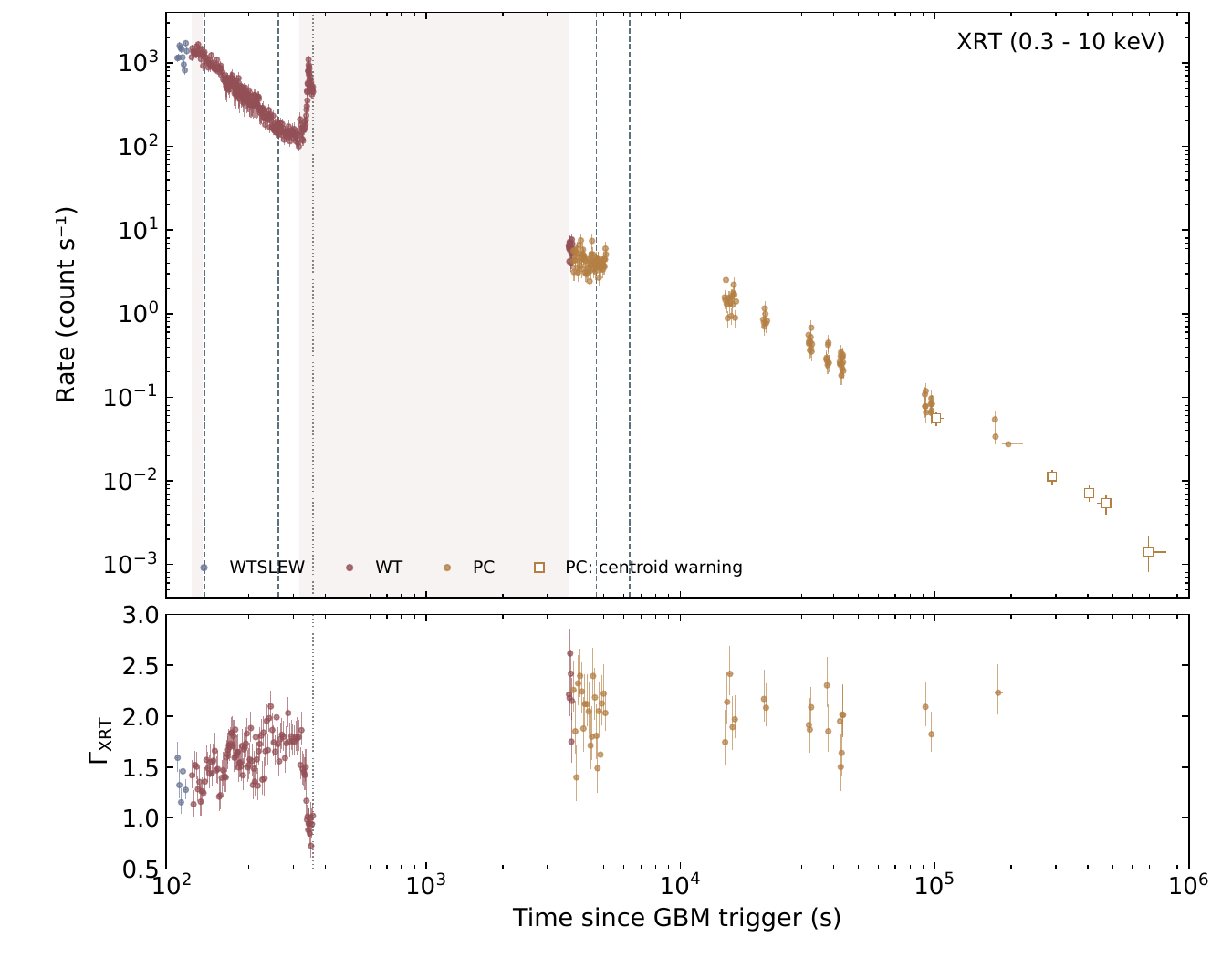}
\caption{Swift/XRT evolution of GRB~260924A from public UKSSDC
products. Top: 0.3--10~keV count rate.
Bottom: photon index inferred from the hardness ratio.
Colours distinguish the observing modes.
Each panel retains its native binning and reported
uncertainties.
Dashed vertical lines and shaded regions indicate
the catalogue break times and excluded intervals
listed in Table~\ref{tab:xrt_catalogue_temporal}; excluded measurements remain plotted.
The dotted line marks the end of the early count-rate
coverage. All times are relative to the GBM trigger.}
\label{fig:XRT_lightcurve}
\end{figure*}

\begin{figure*}
\centering
\includegraphics[width=0.7\linewidth]{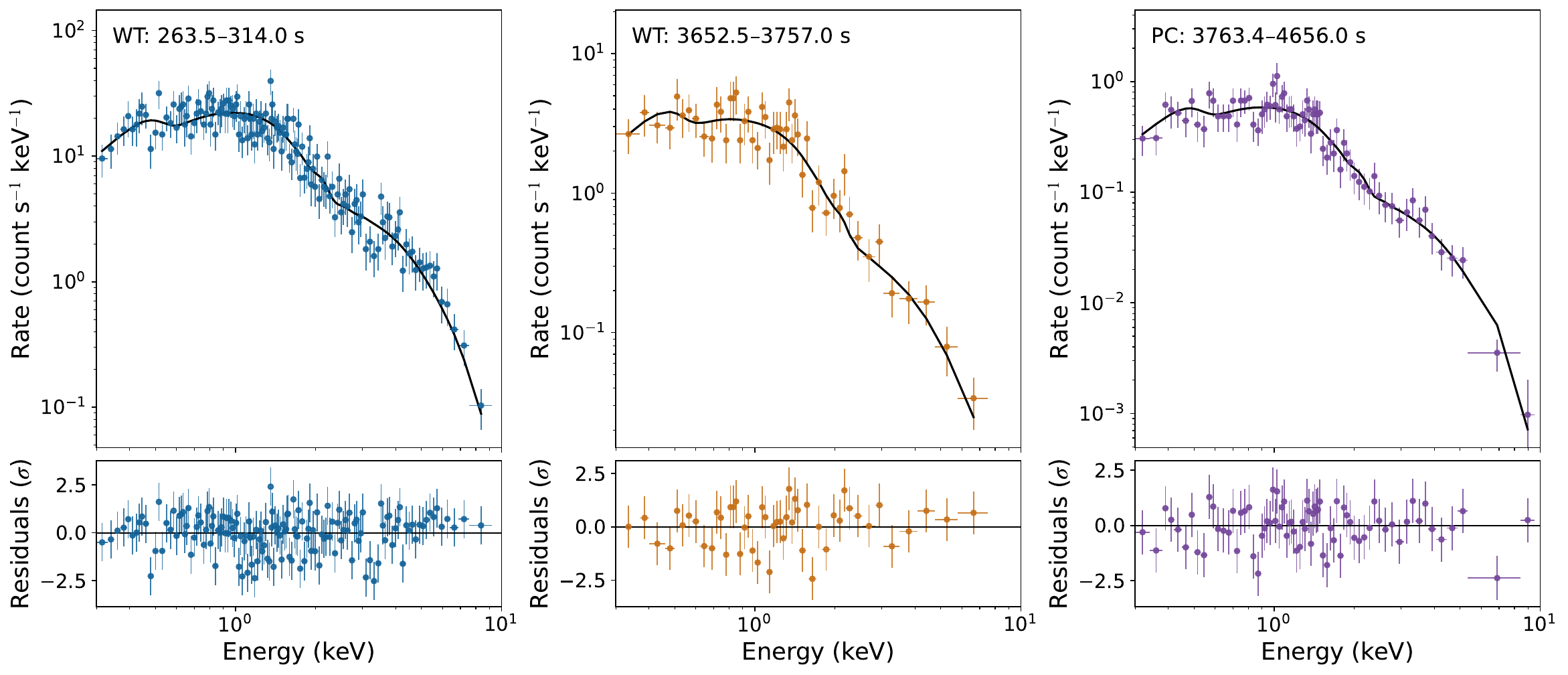}
\caption{Swift/XRT spectra accumulated during 263.5--314.0~s (WT; blue), 3652.5--3757.0~s (WT; orange), and
3763.4--4656.0~s (PC; purple) after the GBM trigger.
Black curves show the joint absorbed power-law fit, folded through the instrumental responses, with a shared intrinsic column density and independent photon indices and normalizations (Table~\ref{tab:xrt_common_absorption}).
Lower panels show normalized residuals, $(D_i-M_i)/\sigma_i$, using the same displayed binning. Horizontal and vertical bars indicate bin widths and $1\sigma$ statistical uncertainties, respectively.
Regrouping was applied only for display.}
\label{fig:XRT_spectra}
\end{figure*}

\begin{figure*}
\centering
\includegraphics[width=0.5\linewidth]{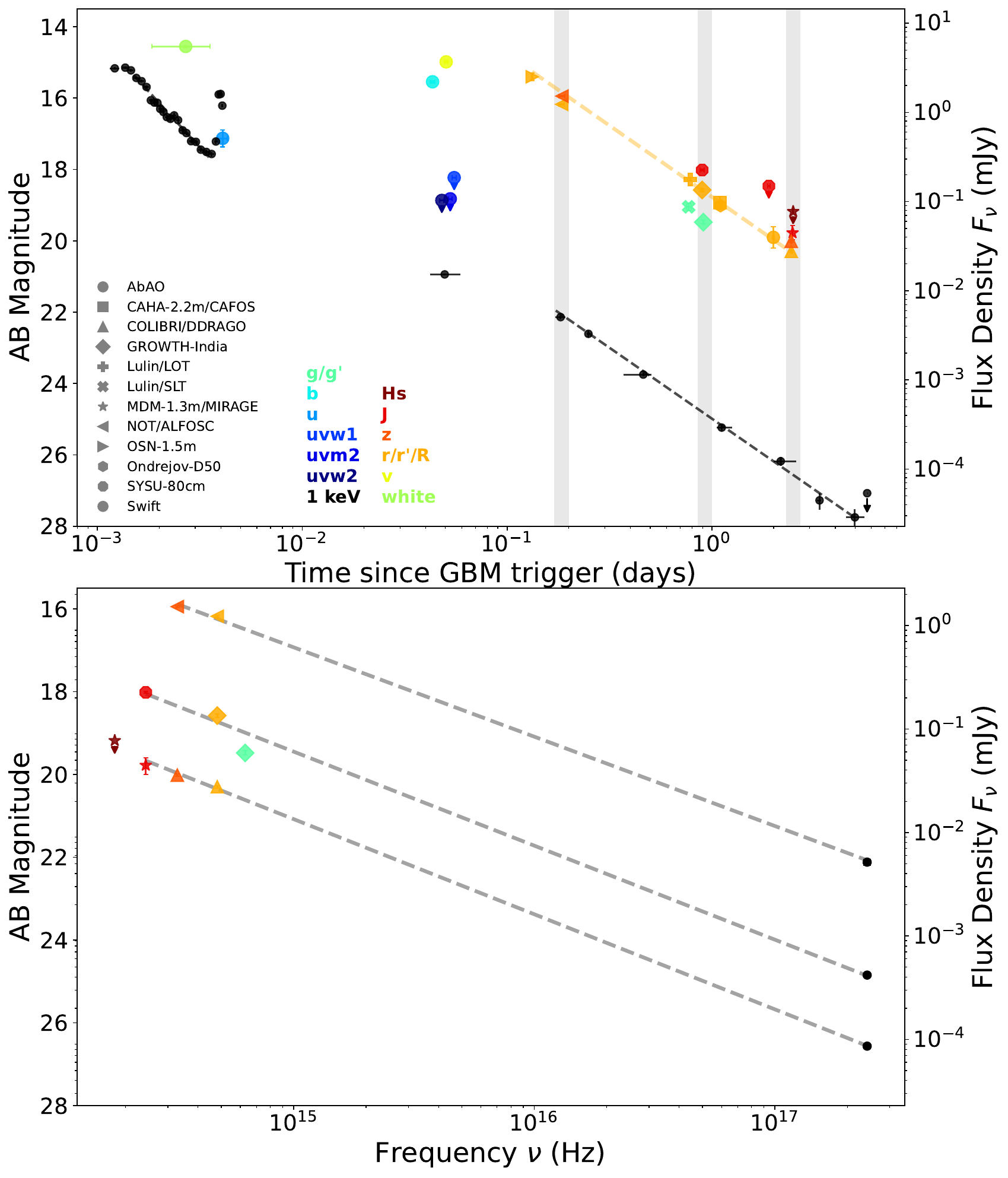}
\caption{
Optical/NIR and X-ray evolution of GRB~260924A.
Top: multiband photometry from the facilities indicated
in the legend and the \textit{Swift}/XRT 1-keV light curve.
Magnitudes are in the AB system and corrected for
Galactic extinction.
The orange dashed line shows the power-law fit
to the $r/r'$ detections.
The early and late black dashed lines show the
catalogue XRT segment (Table~\ref{tab:xrt_catalogue_temporal}) and our temporal fit
used for the SED reconstruction, respectively.
Grey regions mark the SED intervals.
Bottom: optical/NIR-to-X-ray SEDs at
0.18--0.20, 0.85--1.00, and 2.30--2.70~d.
Dashed lines show single-power-law fits;
spectral indices and fit statistics are listed
in Table~\ref{tab:optical_xrt_sed}.
Times are relative to the GBM trigger.
}
\label{fig:lc_sed}
\end{figure*}

\begin{figure*}
\centering
\includegraphics[width=0.7\linewidth]{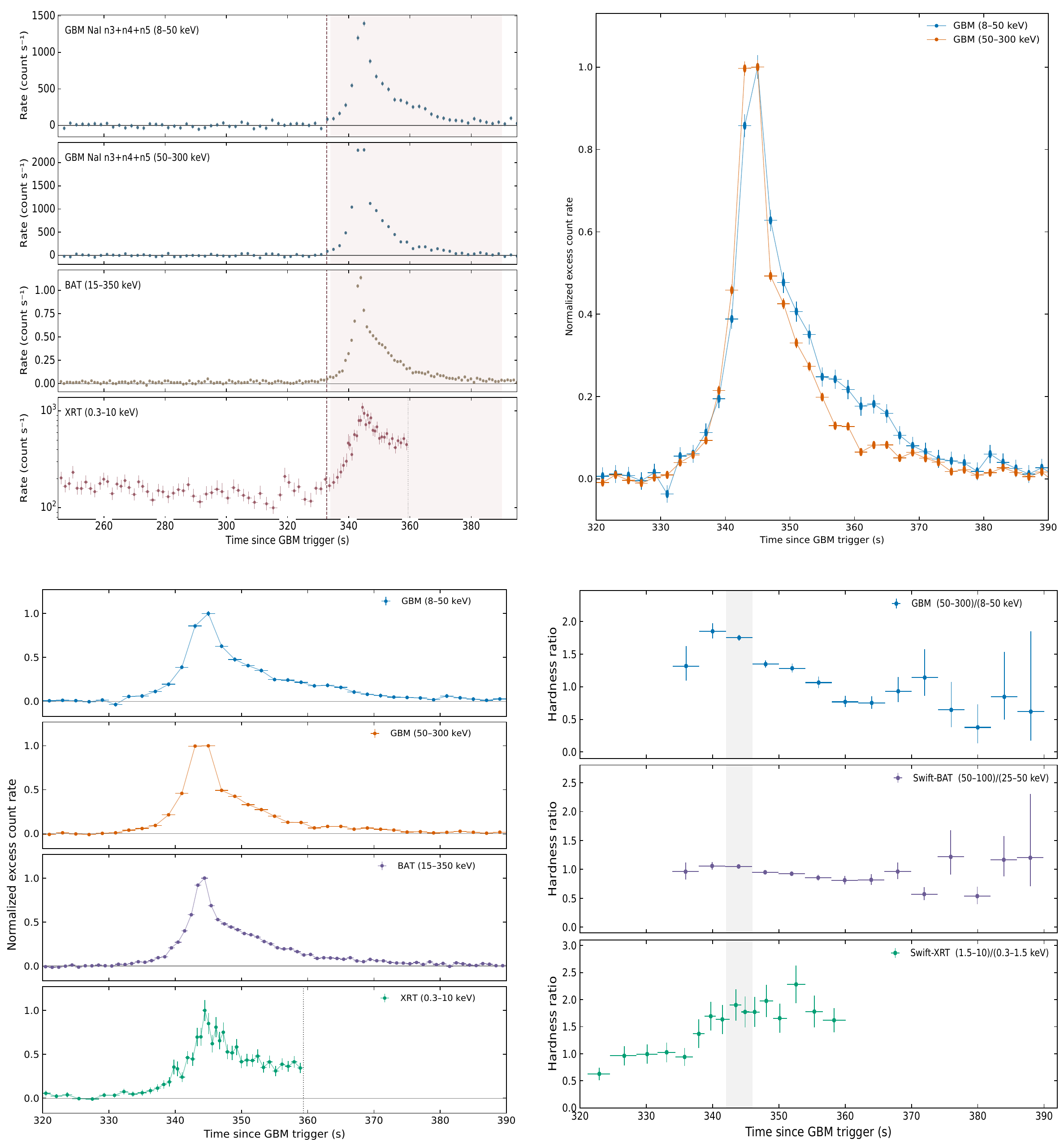}
\caption{Temporal structure and hardness evolution of the
second bright episode in GRB~260924A.
Upper left: GBM, BAT, and XRT count-rate light curves.
The dashed line marks 332.8~s, and the shaded
interval spans 334--390~s.
Upper right: GBM profiles after subtraction of the
local underlying level and normalization to their
respective peak bins.
Lower left: corresponding normalized profiles
for GBM, BAT, and XRT.
Lower right: count-rate hardness ratios in the
indicated bands; shading marks 342--346~s.
Hardness measurements are listed in Table~\ref{tab:second_pulse_hardness}.
Connecting lines guide the eye.
Count-rate and hardness amplitudes are not directly
comparable between instruments.
All times are relative to the GBM trigger. The dotted vertical line in the XRT panels marks
the end of the available coverage.}
\label{fig:second_pulse}
\end{figure*}

\begin{figure*}
\centering
\includegraphics[width=0.65\linewidth]{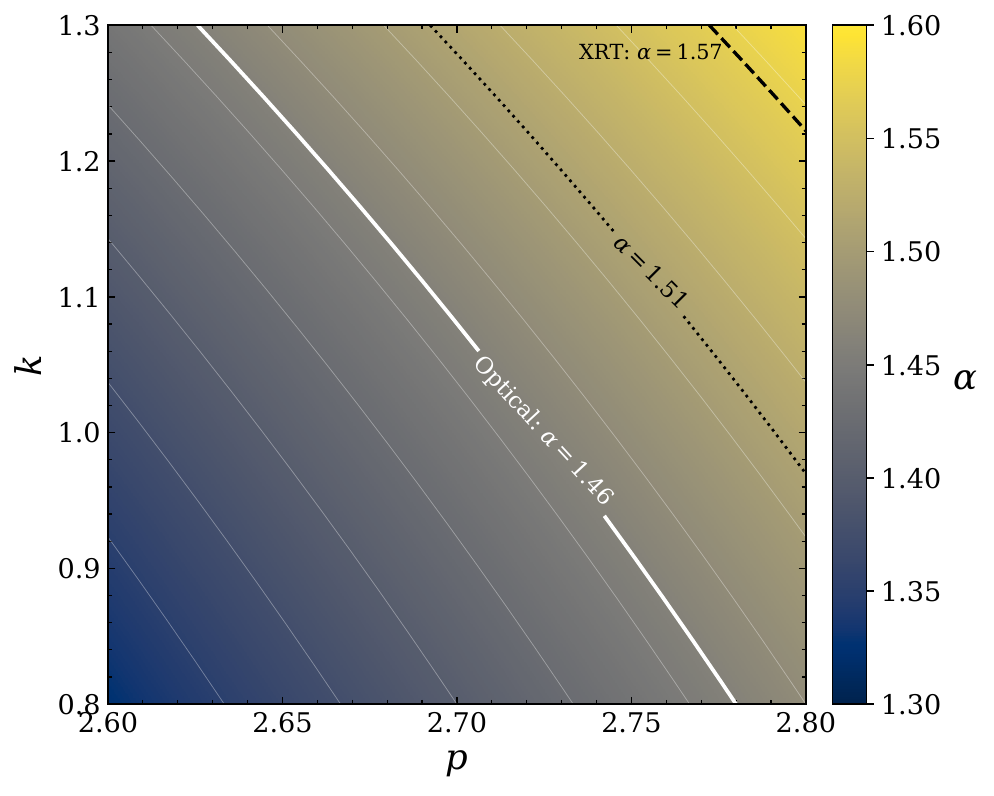}
\caption{Predicted synchrotron temporal decay index \(\alpha\) as a function of the electron index \(p\) and external density gradient \(k\), for \(n(R)\propto R^{-k}\) and \(\nu^{\rm syn}_m<\nu_{\rm opt}<\nu_X<\nu^{\rm syn}_c\). The calculation assumes an adiabatic blast wave before a jet break, with constant energy and microphysical parameters. The colour scale represents \(\alpha^{\rm syn}=3(p-1)/4+k/[2(4-k)]\). The white solid and black dashed contours mark the
representative optical and X-ray decay indices,
$\alpha=1.46$ and $1.57$, respectively; the dotted
contour indicates $\alpha=1.51$. The contours are not statistical confidence regions.}
\label{fig:optical_xrt_closure}
\end{figure*}

\begin{figure*}
\centering
\includegraphics[width=\textwidth]
{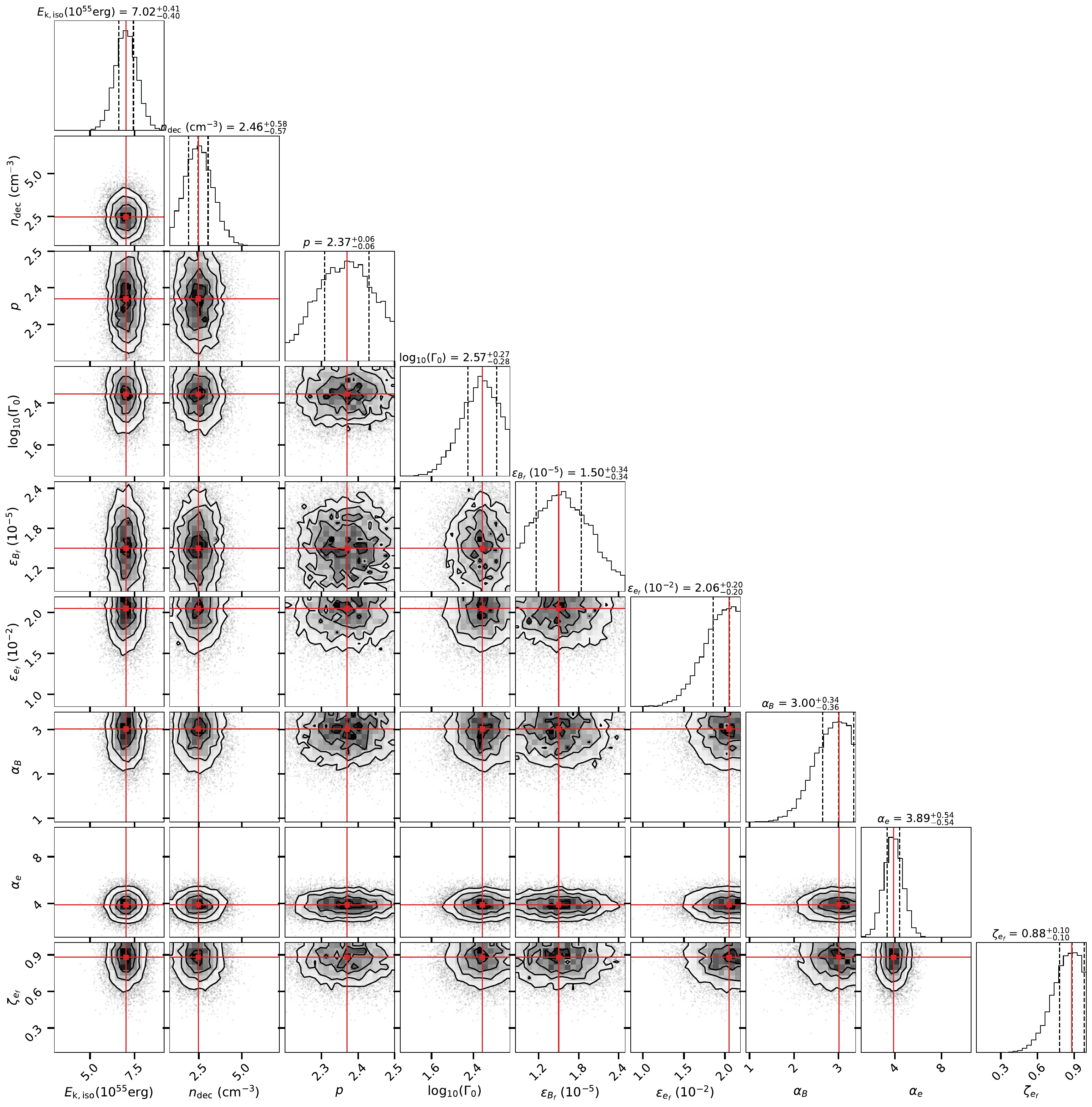}
\caption{Posterior distributions for the forward-shock
model summarized in Table~\ref{tab5:fit}. Diagonal panels show
the marginalized one-dimensional distributions;
off-diagonal panels show the joint distributions
of parameter pairs.}
\label{fig:mcmc_fs}
\end{figure*}

\begin{figure*}
\centering
\includegraphics[width=0.65\textwidth, height=0.3\textheight]{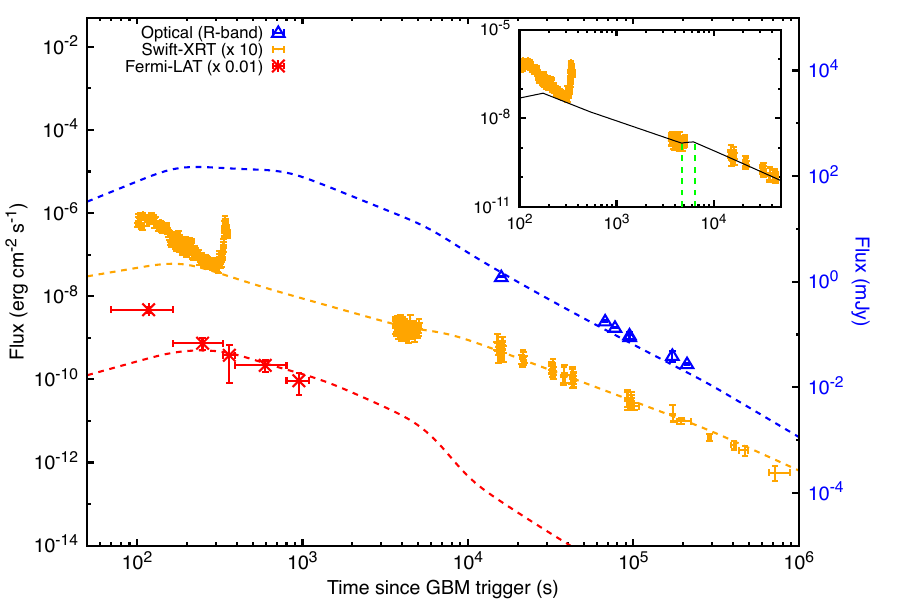}
\includegraphics[width=0.65\textwidth, height=0.3\textheight]{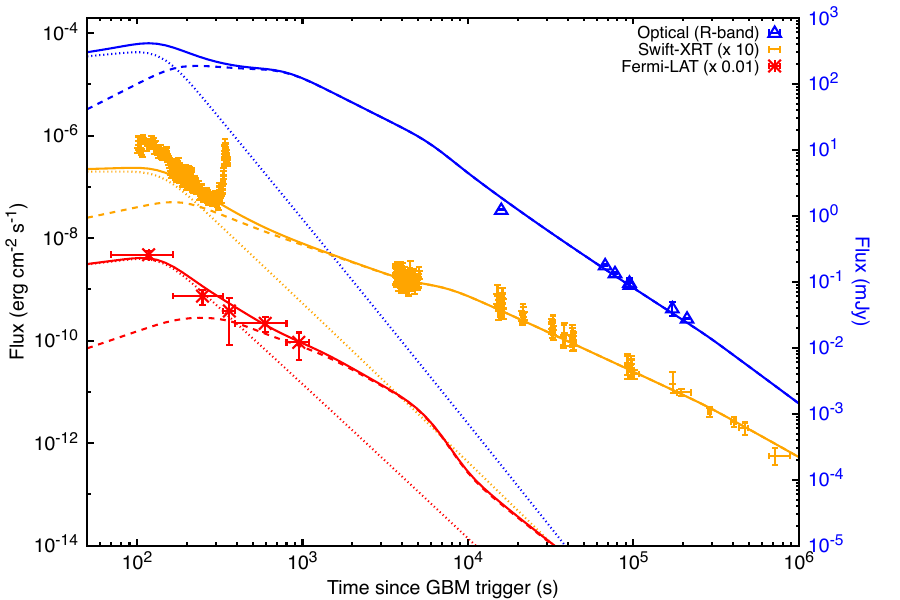}
\caption{Multiwavelength observations and model light curves
for the forward-shock model (top) and the combined
forward- and reverse-shock model (bottom).
The upper inset shows the XRT light curve without
smoothing around the rebrightening. In the lower panel, dashed and dotted curves show the forward- and reverse-shock contributions,
respectively; solid curves show their sum. X-ray and LAT fluxes are scaled as indicated in the legends. Optical flux densities refer to the right-hand axes. The model curves use the posterior median
parameters listed in Table~\ref{tab5:fit}.}
\label{fig:multi_afterglow}
\end{figure*}

\begin{figure*}
    \centering
    \includegraphics[width=0.7\linewidth, height=0.35\textheight]{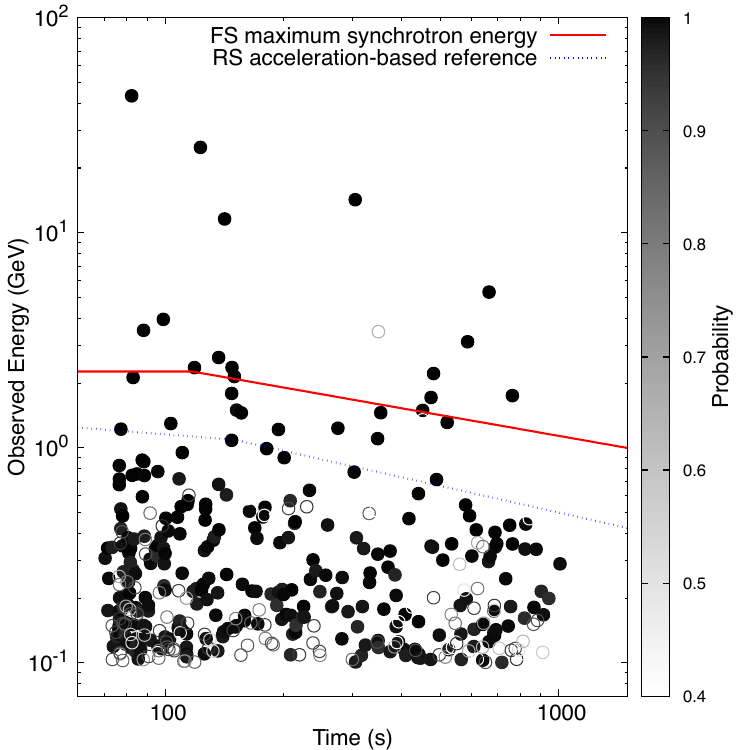}
\caption{ LAT photon energies versus arrival time for
GRB~260924A in the observer frame, with times
relative to the GBM trigger. Greyscale indicates model-dependent association probabilities; darker symbols denote higher values.
The red solid curve shows the forward-shock maximum synchrotron energy, and the blue dotted curve shows the reverse-shock acceleration-based reference. After shock crossing, the latter does not track
the evolving cutoff of the cooling electrons or delayed high-latitude emission. Both curves use the posterior median parameters of the combined shock model in Table~\ref{tab5:fit}.
}
\label{fig:synchrotron_limits}
\end{figure*}

\begin{figure*}
\centering
\includegraphics[width=0.7\linewidth, height=0.35\textheight]{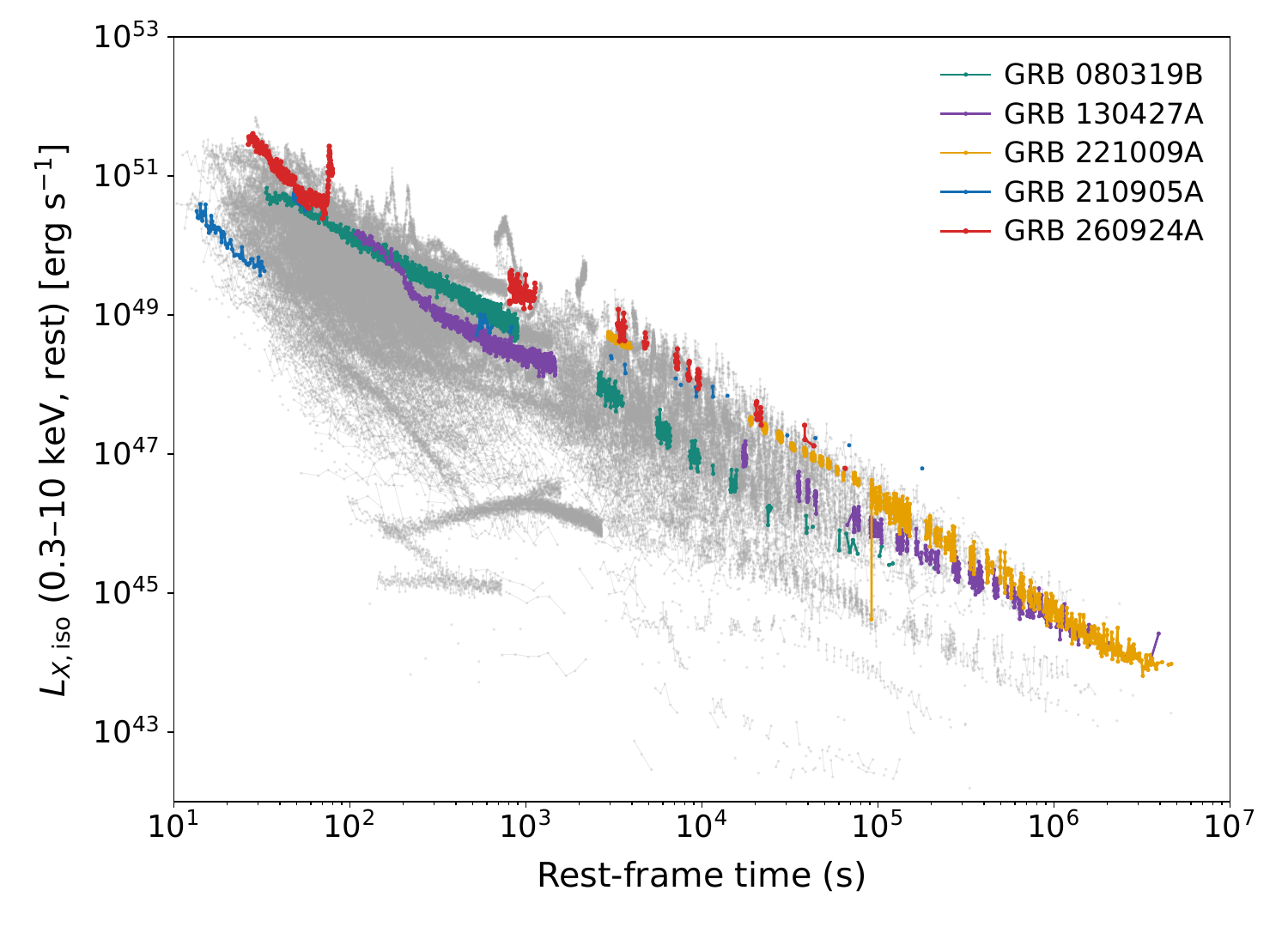}
\caption{Rest-frame 0.3--10~keV isotropic-equivalent X-ray
luminosity light curves of GRB~260924A (red)
and the comparison bursts identified in the legend.
Grey curves show 410 additional \textit{Swift}/XRT GRBs
with published redshifts and usable repository products.
Luminosities are absorption-corrected and transformed
to the rest-frame band using power-law spectra
and the cosmology specified in Section~\ref{sec:cons_phys_parameter}.
For GRB~260924A, interval-specific spectral conversions
are used where available; mode-averaged conversions
are retained elsewhere, including the flare.
Times are divided by $(1+z)$, using the GBM trigger
for GRB~260924A and repository time origins for
the other events, including GRB~221009A.
Only detections are shown; flares are retained.
XRT products are from the UK\textit{Swift}Science Data Centre
\citep{Evans2009}.}
\label{fig:x-ray_luminosity}
\end{figure*}

\begin{figure*}
\centering
\includegraphics[width=0.7\linewidth, height=0.35\textheight]{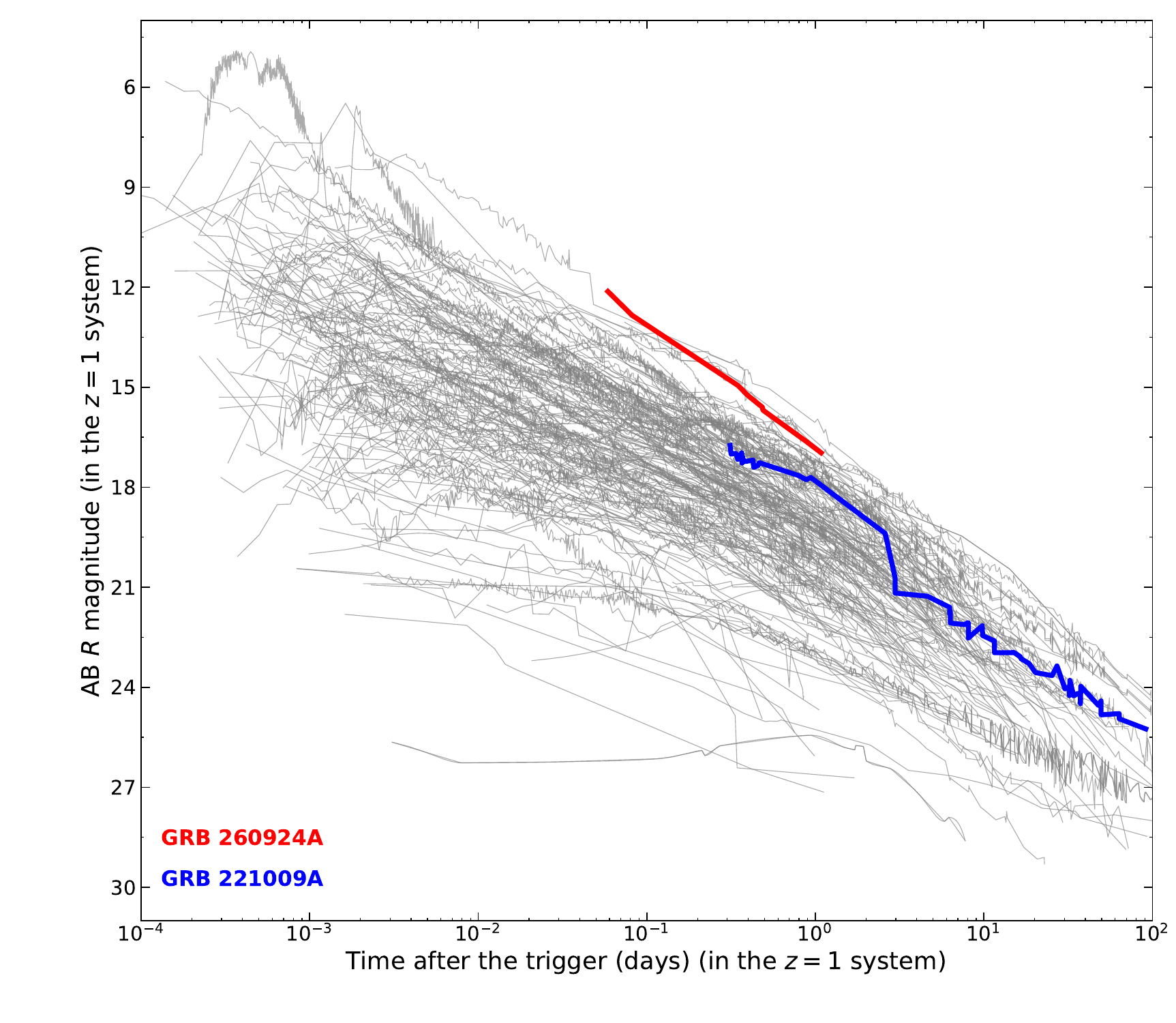}
\caption{Optical afterglow of GRB~260924A (red), compared
with the long-GRB sample of \citet{Kann2006,Kann2010, Kann2011,Kann2024} and GRB~221009A
from \citet{OConnor2023}. Light curves are shifted to a common redshift
of $z=1$ and expressed as AB $R$-band magnitudes,
accounting for luminosity distance, time dilation,
Galactic extinction, and $k$- and colour corrections.
For GRB~260924A, the transformation uses
$z=3.51$ and $\beta=0.9$.}
\label{fig:kann}
\end{figure*}

\begin{figure*}
    \centering
    \includegraphics[width=0.7\linewidth, height=0.35\textheight]{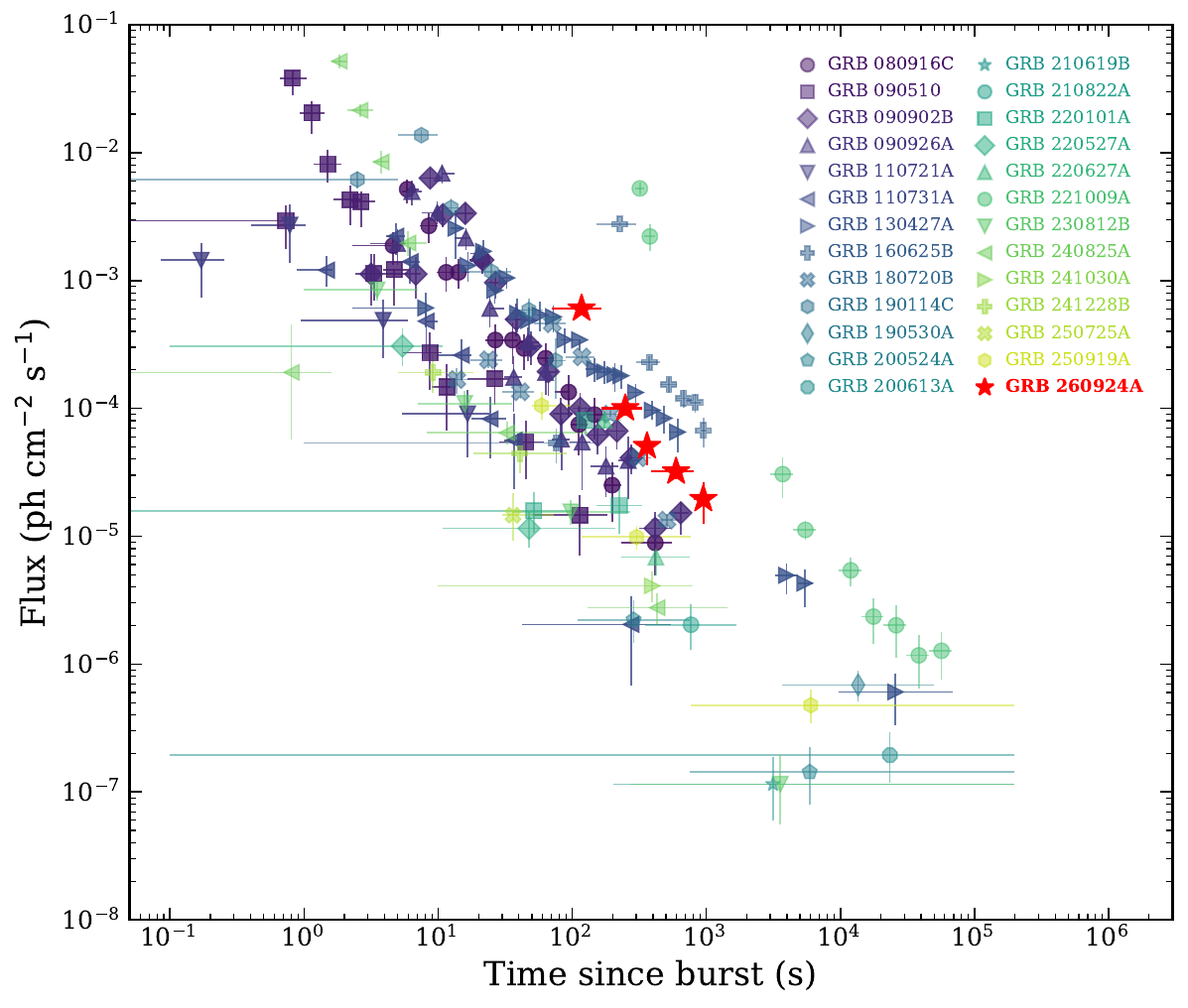}
\caption{Comparison of the \textit{Fermi}/LAT photon-flux light curve
of GRB~260924A (red stars) with a selected sample of bright
LAT-detected GRBs. Published comparison light curves were
taken from \citet{2013ApJ...763...71A, 2013ApJS..209...11A, 2014Sci...343...42A, 2019ApJ...885...29F, 2019ApJ...879L..26F}. The light curves of GRBs detected
from 2020 onwards, including GRB~260924A, were derived
in this work for this comparison. Times are measured
in the observer frame relative to the respective burst
triggers. Photon fluxes are integrated above 100~MeV,
with upper energy limits of 100, 300, or
1000~GeV depending on the dataset;
GRB~260924A uses the 0.1--1000~GeV band.}
\label{fig:LAT_comp}
\end{figure*}

% Tables and figures placed here
\clearpage

\appendix

\section{Additional details of the GBM spectral analysis}
\label{app:gbm_systematics}

This appendix provides additional details of the response construction, background estimation, exposure calculation, and sensitivity tests used in the \textit{Fermi}/GBM spectral analysis described in Section~\ref{sec:gbm_analysis}.

\subsection{Effective response construction}
\label{app:gbm_response}

For each detector, we constructed an effective response by averaging the time-dependent RSP2 response over the corresponding spectral extraction interval. The response at the centre of each temporal bin was obtained by linear interpolation between the temporal centres of adjacent response matrices and weighted by the total observed counts in that bin, including both source and background counts.

For the first-episode spectra covering $-6.1$--50, $-6.1$--164.9, 50--115, and 115--164.9~s, the response weights were derived from the total observed counts in each detector's fitted energy range, using 1.024-s bins. The weighting included the full bins intersecting each extraction interval, so their coverage could extend slightly beyond the spectral boundaries. For the two subdivisions of the $-6.1$--50~s
interval, the weighting bins were instead clipped to the extraction boundaries.

For the second bright episode, the effective response was constructed separately for each detector. The weighting grid had a nominal spacing of 1.024~s and included the exact extraction boundaries and the response transition at $T_0+366.597954$~s. Counts were extracted directly from the TTE data using the full detector energy range adopted before applying the channel exclusions used in the principal spectral fit. This procedure combined RSP5 and RSP6 using detector-specific weights.

As a sensitivity test, we replaced the total-count weighting with weighting by dead-time-corrected exposure. We also compared the linearly interpolated responses with piecewise responses assigned according to their nominal validity intervals. The resulting spectral parameters were consistent with those obtained using the adopted response prescription.

\subsection{Background estimation and uncertainty propagation}
\label{app:gbm_background}

For the first-episode spectra, the background
was fitted independently in each detector channel
using first- and second-order polynomials on
1.024-s temporal bins over the nominal off-source
intervals $[-125,-35]$ and $[250,320]$~s
relative to $T_0$.   The fits used the two-pass weighted least-squares
procedure implemented in the Gamma-ray Data Tools
(GDT). Interval-averaged background rates and their uncertainties were
computed from the fitted coefficients and
their full covariance matrix.

The two time-resolved subdivisions of the initial episode used the same off-source intervals. Their background estimates therefore share the same off-source data and have correlated background uncertainties. These correlations are relevant when comparing the two subintervals, although each spectrum was fitted independently.

For the second bright episode (332.8--390~s),
the off-source intervals were 176--200,
224--240, and 410--476~s for the NaI detectors,
and 190--240 and 410--476~s for b0, relative
to $T_0$. The principal analysis used linear
background models fitted independently in each
channel by maximizing the Poisson likelihood
on 2-s bins. Positivity was imposed on a grid
spanning 176--480~s; this grid was not used
as a continuous off-source interval.
The coefficient covariance matrix was estimated
from the inverse Fisher information matrix
and propagated to the integrated background
counts, accounting for the live exposure.

The principal spectral fit retained 391 channels:
119 from n3, 119 from n4, 90 from n5, and
63 from b0, spanning 8.73--884.51,
8.82--899.28, 52.19--885.48, and
318.35--4880.69~keV, respectively, with no
internal channel exclusions within these ranges.

The resulting background estimates and variances supplied the Gaussian background constraints used in the spectral likelihood. The temporal polynomial coefficients were not refitted during spectral fitting. Correlations between detector channels and uncertainty associated with the choice of polynomial order were not included in these variances. We therefore additionally examined alternative polynomial orders as part of the sensitivity tests described in Section~\ref{app:gbm_sensitivity}.

\subsection{Exposure calculation and time-resolved extraction}
\label{app:gbm_exposure}

For the $-6.1$--50, $-6.1$--164.9, 50--115,
and 115--164.9~s spectra, we used the
detector-specific live exposures returned by
GDT in the extracted PHA products.
These exposures were also used to convert the
estimated background rates and their uncertainties
into counts.

For the diagnostic subdivision of the initial bright episode, a single temporal boundary at $T_0+12.288$~s was selected from the 0.512-s GBM total-count light curve before fitting the two spectra. The boundary was chosen by visual inspection
to separate the initial bright double-peaked
complex and its rapid decline near 12--13~s
from the subsequent weaker emission.
No alternative cuts were tested.
The quoted precision identifies the selected
bin edge, rather than an uncertainty on the
physical end of the pulse. Counts and live exposures were extracted independently for the two resulting intervals, while retaining the same spectral channel selection and off-source background intervals used for the integrated spectrum.

For each subinterval, the live exposure was calculated from the extraction duration after accounting for detector dead time from all channels. This convention differs by at most 0.163 per cent from the exposure recorded in the archived PHA products and has a negligible effect on the spectral results.

\subsection{Poisson--Gaussian likelihood}
\label{app:gbm_likelihood}

The auxiliary background parameter was profiled
analytically in each channel without a
non-negativity constraint. The total Poisson
mean was constrained to be non-negative and
strictly positive in channels containing
observed counts.

\subsection{Blackbody implementation}
\label{app:gbm_blackbody}

We describe the BB component with a Planck photon
spectrum \citep{Ryde2004,RydePeer2009}
\begin{eqnarray}
\label{eq:gbm_blackbody}
N_{\rm BB}(E)=
\frac{A_{\rm BB}E^2}{\exp(E/k_{\rm B}T)-1},
%\label{eq:blackbody}
\end{eqnarray}
where $E$ is the photon energy, $T$ is the blackbody temperature, and $k_{\rm B}$ is Boltzmann's constant. Both $E$ and $k_{\rm B}T$ are expressed in keV, and $N_{\rm BB}(E)$ is the differential photon flux in $\mathrm{ph\,cm^{-2}\,s^{-1}\,keV^{-1}}$. For the numerical implementation, we express its normalization as $A_{\rm BB}=10^{-6}A_6\,{\rm ph\,cm^{-2}\,s^{-1}\,keV^{-3}}$, where $A_6$ is dimensionless.  We imposed $A_6\geq0$ and restricted the temperature to $5\leq k_{\rm B}T\leq300$~keV. To test sensitivity to the allowed normalization range, we repeated the fits using upper bounds of $A_6=20$, 40, and 80.

\subsection{Sensitivity tests}
\label{app:gbm_sensitivity}

In addition to the response tests described in
Appendix~A1, we examined the sensitivity of the
spectral results to linear and quadratic temporal
background models and to the detector-channel
selections used in the fits.

For the second bright episode, particular attention was given to the stability of the results under alternative background descriptions and channel selections because of the weaker signal and the more complex temporal environment surrounding this episode. The $T_0+250$ to $T_0+320$~s interval was not used to estimate the background for this episode.

As an additional background check, we repeated
the Band and Band+BB fits over 62--68~s, excluding
250--320~s from the background selection.
The alternative windows were $[-125,-35]$ and
$[176,200]$~s for NaI, and $[-125,-35]$ and
$[190,200]$~s for BGO, relative to the GBM trigger.
The fit improvement persisted, with
$\Delta{\rm PG}=27.13$ and $34.25$ for linear
and quadratic backgrounds, respectively, and
$k_{\rm B}T\simeq41$--$42$~keV.
This supports the persistence of the candidate
under the tested background selections, although
temporal background residuals remain.

\section{Additional details of the LAT likelihood analysis}
\label{app:lat_systematics}

This appendix provides additional details of the likelihood implementation, source-model configuration, uncertainty propagation, and sensitivity tests used in the \textit{Fermi}/LAT analysis described in Section~\ref{sec:lat_analysis}.

\subsection{Likelihood implementation and source model}
\label{app:lat_likelihood}

The analysis used the \texttt{UnbinnedObs}
and \texttt{UnbinnedAnalysis} interfaces
with the \texttt{NewMinuit} optimizer and
the source configuration described in
Section~2.2. The GRB spectrum was implemented
with \texttt{PowerLaw2} over 0.1--1000~GeV.
The archived model-building script specified
the catalogue file \texttt{gll\_psc\_v35.fit}.\footnote{The original catalogue FITS file and its resolved input path were not available in the audited analysis package; the catalogue identification therefore refers to the archived script configuration.}

\subsection{Test statistic and diffuse-normalization bounds}
\label{app:lat_ts}

We calculated $\mathrm{TS}=2(\ln\widehat{\mathcal L}_1-
\ln\widehat{\mathcal L}_0)$ for models with and without the GRB,
reoptimizing both diffuse normalizations within $[0,30]$ under each hypothesis. Sensitivity to these bounds is examined in Appendix~B4.

The reported photon-flux and photon-index errors are nominal $1\sigma$ covariance uncertainties from the source-present fit;
they exclude instrumental and background-model systematics.
\subsection{Energy-flux calculation and uncertainty propagation}
\label{app:lat_flux}

Energy fluxes were obtained by integrating the archived \texttt{PowerLaw2} spectra over 0.1--1000~GeV. Flux uncertainties were
propagated from the full-fit covariance matrix, retaining the normalization--index covariance with the diffuse normalizations free during fitting.

For Figure~\ref{fig:LAT_lightcurve}, photon and energy fluxes were also evaluated over 0.1--300~GeV using the same covariance treatment. This display
conversion did not modify the original 0.1--1000~GeV likelihood fits.

\subsection{Sensitivity to the background model}
\label{app:lat_sensitivity}

We examined the sensitivity of the LAT results to assumptions entering the background model and likelihood optimization. In particular, we repeated the relevant likelihood calculations using wider bounds on the Galactic and isotropic diffuse normalizations. This test was especially important for the first principal interval, for which the inferred source significance showed the greatest sensitivity to the allowed diffuse-normalization range.

The catalogue-source configuration, event selection, GRB spectral model, and energy range were otherwise kept unchanged when evaluating the effect of the diffuse-normalization bounds. The resulting variations were used as a diagnostic of background-model sensitivity rather than incorporated into the formal statistical uncertainties quoted for the principal fits.

Accordingly, the reported TS values and covariance-based parameter uncertainties should be interpreted as conditional on the adopted source and diffuse-background model. No additional systematic uncertainty associated with alternative Galactic diffuse templates, instrumental response functions, or catalogue-source configurations was included.

For the five integrated intervals, the principal $[0,30]$ diffuse-bound TS values are 1058.36, 282.41, 63.55, 309.28, and 26.95. Under wider bounds, the corresponding stable values are 863.90, 281.62, 63.55, 309.27, and 26.95. The primary table retains the common $[0,30]$ convention. The larger differences in the first interval mainly reflect the improved source-removed hypothesis under wider bounds. Stability against increasing a parameter bound is not a test of the adequacy of the background templates.

\subsection{Updated temporal fit with fixed normalized exposure weights}
\label{app:lat_temporal_weights}
We fitted the final interval-averaged photon fluxes using independent Gaussian errors. For each time cell $j$ in interval $i$, the archived integrated response $w_{ij}$ defines a normalized weight. The prediction is
\begin{equation}
\overline F_i=\sum_j\frac{w_{ij}}{\sum_jw_{ij}}\frac{1}{b_{ij}-a_{ij}}\int_{a_{ij}}^{b_{ij}} A\left(\frac{t}{164.9\,\mathrm{s}}\right)^{-\alpha}dt.
\end{equation}
The response is treated as constant within each cell. The primary calculation retains the archived refined weights for the first four intervals and the 16-cell \texttt{gtexposure} representation for the last interval. All cell boundaries match the final good-time intervals. The final spectral indices differ from those used to generate the weights by less than $9\times10^{-5}$ in the first four intervals and by 0.00436 in the last. We retain the original normalized weights rather than claiming exposures regenerated at the final indices. Their sensitivity to this small index change has not been separately propagated.

Replacing the last-interval weights by the archived 16-cell $N_{\rm pred}/F$ weights changes the fitted slope from 1.2980 to 1.2969 and the first-bin excess ratio from 2.2487 to 2.2512. This comparison tests the effect of the previously audited difference in normalized temporal shape; it does not resolve the absolute response discrepancy between the two methods. The normalization $A$ is fitted freely. The first bin is excluded from fitting and used only to assess the extrapolation; the 250--320~s diagnostic and overlapping subdivisions are not additional independent measurements. The second-pulse interval remains included, so this fit does not isolate a pure afterglow contribution.

The covariance uses the absolute quoted flux errors without rescaling by reduced $\chi^2$. The asymmetric errors on $A$ and $\alpha$ are one-parameter $\Delta\chi^2=1$ profile intervals of this Gaussian flux fit, not LAT event-likelihood profiles. The extrapolation and excess-ratio errors use local covariance propagation. Instrumental, spectral, background-model, and exposure-weight uncertainties are not included. Analytical cell averages were checked against independent numerical quadrature.

\section{Additional details of the BAT spectral analysis}
\label{app:bat_systematics}

%\subsection{Channel-dependent systematic uncertainties}
%\label{app:bat_syserr}

As a reproducibility check, we subsequently applied the explicitly identified calibration file \texttt{swbsyserr20030101v003.fits} (version 3) to copies of the three fitted spectra using \texttt{batphasyserr} version 1.6 under HEASoft 6.37.1. The resulting \texttt{SYS\_ERR} values matched those preserved in the fitted spectra exactly in all 80 channels of each spectrum.

This comparison provides a retrospective verification of the systematic uncertainties applied in the spectral analysis. The calibration file selected during the original processing was not recorded; consequently, the identification of \texttt{swbsyserr20030101v003.fits} refers to the calibration file used for this subsequent reproducibility check rather than to a contemporaneous record of the original processing.

For the 70-165~s and 334-390~s intervals,
refining the response construction from segments
of at most 5~s to at most 2.5~s changed the fitted
photon indices and energy fluxes by substantially
less than their statistical uncertainties.
The 250-320~s interval retained a single response
and was not subdivided in this test.

\section{Additional details of the optical/NIR and XRT analysis}
\label{app:optical_xrt}

The UV, optical, and NIR photometry was placed on the AB magnitude system and corrected for Galactic foreground extinction adopting $E(B-V)=0.041$ and $R_V=3.1$. For each photometric band, the extinction correction was evaluated at the adopted effective or pivot wavelength of the corresponding filter using the extinction prescriptions described in Section~\ref{sec:optical_observations}.

The integrated 0.3--10~keV XRT energy fluxes
were converted to flux densities at 1~keV
using power-law spectra with
$\Gamma_{\rm WT}=1.49$ and
$\Gamma_{\rm PC}=2.04$, normalized to the
measured band-integrated flux in each mode.

For visualization, the densely sampled XRT light curve was rebinned in time. At $t<0.01$~d, consecutive groups of measurements were combined, while at later times the observations were grouped into predefined temporal intervals. Within each bin, the flux was calculated using inverse-variance weighting, and the representative time was determined from the error-weighted geometric mean of the individual measurement times. The horizontal uncertainties span the temporal coverage of the measurements contributing to each bin. This rebinning was used only for visualization; all temporal fits were performed using the original XRT measurements.

For the SED reconstruction, the geometric mean of the boundaries of each temporal interval was adopted as the representative epoch. Optical/NIR measurements were shifted to the corresponding epoch using the adopted optical temporal decay, with its shared uncertainty propagated into the covariance. The reconstructed SEDs include NOT $r$ and $z$ measurements in the first epoch, GROWTH $r'$ and SYSU $J$ in the second, and COLIBRÍ $r$ and $z$ together with MDM $J$ in the third. Photometric upper limits were displayed but were not included in the spectral fits.

Because an XRT measurement was not necessarily available at each representative epoch, the 1~keV X-ray flux density was evaluated at that epoch using the best-fitting late-time XRT temporal power law. This procedure places the X-ray and optical/NIR measurements at a common characteristic epoch rather than averaging the X-ray emission over a different temporal interval.

\subsection{Test for a late-time X-ray break}
\label{app:xrt_late_break_test}

We used the public \textit{Swift}/XRT PC-mode light curves
provided by UKSSDC \citep{Evans2009,Evans2010}. This independent
check uses the native unabsorbed 0.3--10 keV light curves,
whose count-to-flux conversion tracks the hardness ratio,
rather than the fixed-index 1-keV reconstruction described
above. Times retain the adopted GBM reference,
$t_{\mathrm{GBM}}=t_{\mathrm{Swift}}+4.71$ s.
We selected detection bins entirely later than $10^4$ s,
with no upper time cut. The selections excluding and
including exposures flagged for unreliable centroiding
contain 69 and 72 detections, respectively, with bin edges
spanning $1.49\times10^4$--$2.94\times10^5$ s and
$1.49\times10^4$--$8.80\times10^5$ s.
The products regroup surviving events, so their difference
is not equivalent to removing three bins. The flags indicate
uncertain centroiding, not necessarily erroneous measurements;
we therefore report both selections.

\begin{figure*}
\centering
\includegraphics[width=0.9\textwidth, height=0.35\textheight]
{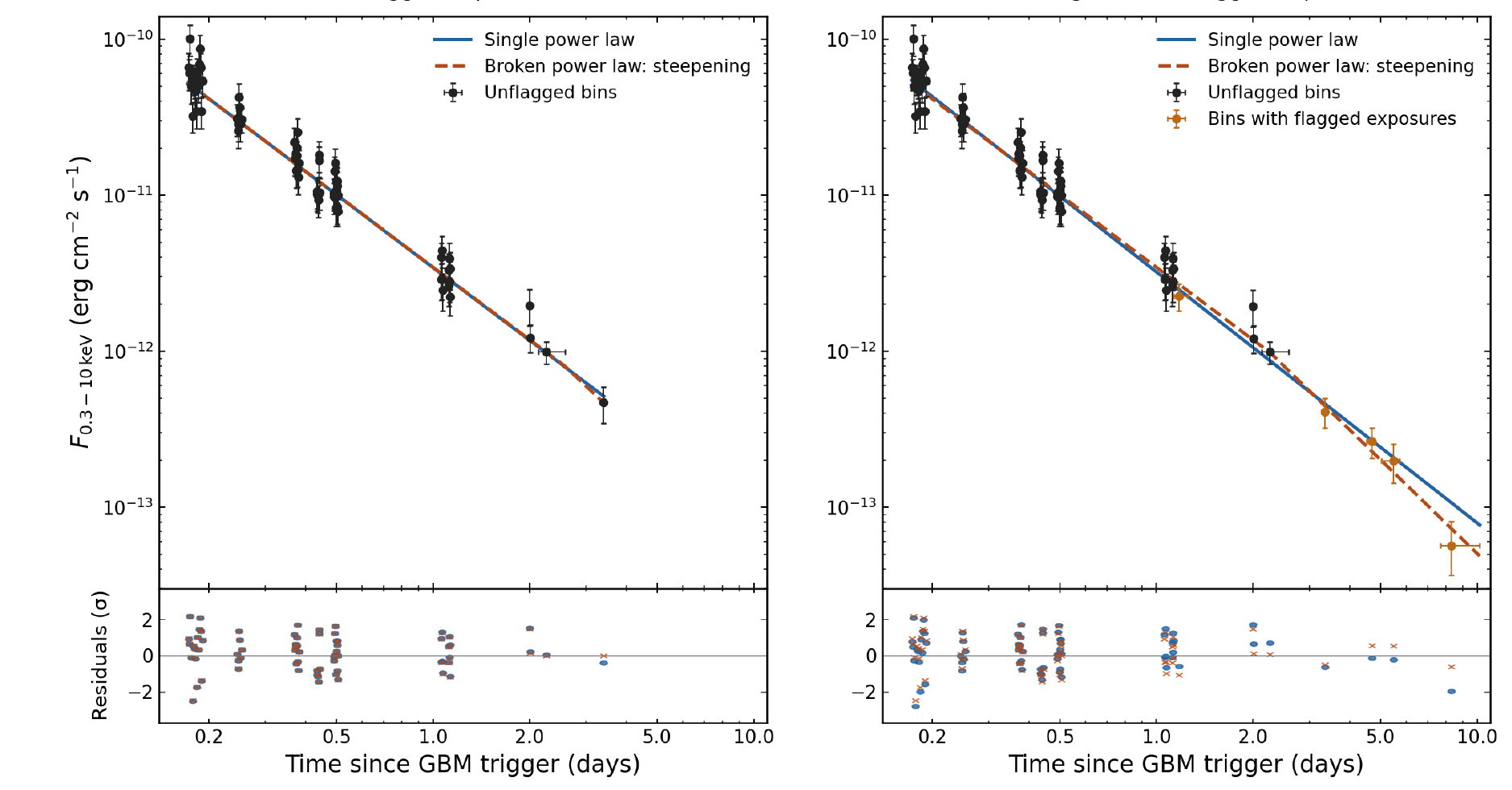}
\caption{
Late-time \textit{Swift}/XRT light curve in the unabsorbed
0.3--10~keV band. Left and right columns exclude and include exposures
flagged for unreliable centroiding, respectively. Top: original detection bins with single-power-law (solid blue) and continuous broken-power-law (dashed orange) fits, the latter constrained to steepen. Orange points identify bins containing flagged exposures.
Curves show instantaneous models, while fits use GTI-averaged predictions.
Bottom: normalized residuals, $(F_i-F_{i,\mathrm{model}})/\sigma_i$,
using the asymmetric-error prescription described in the text; circles and crosses correspond to the single and broken power laws, respectively. Sample sizes and fit statistics are listed in the first two rows of Table~\ref{tab:xrt_late_break_test}.
}
\label{fig:xrt_late_break_test}
\label{lastpage}
\end{figure*}

We compared a single power law with a continuous
broken power law having free normalization,
pre- and post-break slopes, and break time.
Both models used the normalization time $t_{\rm ref}=10^5$~s. We examined both unrestricted slope changes
and the steepening alternative $\alpha_2\geq\alpha_1$;
Table~\ref{tab:xrt_late_break_test} and
Fig.~\ref{fig:xrt_late_break_test} show the latter.
We searched $t_{\mathrm{br}}$ between the first and last
detection centres using profiles and multiple starting
points. No break time or slope was imposed from another
waveband or from the physical afterglow model.

Predictions were averaged over the intersections of each
bin with its actual good-time intervals (GTIs), weighting
by GTI duration. Fits minimized squared residuals in linear
flux, using the upper reported uncertainty when the model
exceeded the datum and the lower uncertainty otherwise.
Quoted slope errors correspond to nominal profile intervals
with $\Delta\chi^2=1$. Count-to-flux conversion uncertainties
and covariances are not included. Controls in count rate
and in the native observed 1-keV flux density, as well as
alternative bin-averaging prescriptions, gave the same
qualitative conclusion. These products are correlated
and their statistics were not combined.

The unflagged band-flux selection provides essentially no
improvement for a steepening, and its entire searched
break-time range lies within $\Delta\chi^2=1$ of the minimum.
Including flagged exposures yields a broad minimum near
$t_{\mathrm{br}}=2.26$ d, with $\alpha_1=1.54$ and
$\alpha_2=1.99$; the nominal profile region
$\Delta\chi^2\leq1$ extends approximately from 1.66 to 7.45 d.
This is not a robust measurement of a break epoch.
Allowing either steepening or flattening does not establish
a late break in the unflagged data either. The unflagged
count-rate upper limit near $8.59\times10^5$ s is also
consistent with the single-power-law extrapolation and
was not treated as a Gaussian detection.

We do not assign a detection significance to
$\Delta\chi^2$, because the break time is
unidentified under the null hypothesis and
zero steepening is a parameter boundary
\citep{Protassov2002}.
The hardness ratios do not establish spectral
constancy across the candidate break.
These data neither require a jet break nor
provide a robust break epoch for the physical model.

\begin{table*}
\centering
\caption{
Comparison of single power laws  and continuous broken
power laws constrained to steepen in the late-time
XRT data. The break time is free. ``Excluded'' and
``Included'' refer to exposures flagged for unreliable
centroiding.
$\Delta\chi^2=\chi^2_{\rm PL}-\chi^2_{\rm BPL}$;
the quoted degrees of freedom are nominal
($N-2$ and $N-4$).
Slope uncertainties are nominal profile intervals with
$\Delta\chi^2=1$. GTI-averaged model predictions are used;
flux-conversion uncertainty is not propagated.
The two products are correlated and their statistics
must not be added.
}
\label{tab:xrt_late_break_test}
\small
\setlength{\tabcolsep}{5pt}
\renewcommand{\arraystretch}{1.2}
\begin{tabular}{llcrrrr}
\hline
Product &
Flagged exposures &
$N$ &
$\alpha_{\rm PL}$ &
$\chi^2_{\rm PL}/{\rm dof}$ &
$\chi^2_{\rm BPL}/{\rm dof}$ &
$\Delta\chi^2$ \\
\hline

$F_{0.3-10\,\mathrm{keV}}$ &
Excluded & 69 &
$1.55^{+0.04}_{-0.04}$ &
66.32/67 & 66.15/65 & 0.17 \\

$F_{0.3-10\,\mathrm{keV}}$ &
Included & 72 &
$1.61^{+0.03}_{-0.03}$ &
72.43/70 & 67.02/68 & 5.41 \\

Count rate &
Excluded & 69 &
$1.56^{+0.04}_{-0.04}$ &
62.12/67 & 62.09/65 & 0.03 \\

Count rate &
Included & 72 &
$1.61^{+0.03}_{-0.03}$ &
66.42/70 & 62.64/68 & 3.77 \\

\hline
\end{tabular}
\end{table*}

%\clearpage
%\bsp

\end{document}